\documentclass[12pt]{article}
\usepackage{amsmath,amsfonts,enumerate,array}
\usepackage{graphicx}
\usepackage[usenames]{color}
\usepackage[english]{babel}
\usepackage{fancybox}
\usepackage{chngpage} % allows for temporary adjustment of side margins

\newcommand{\captionfonts}{\small}

\makeatletter  % Allow the use of @ in command names
\long\def\@makecaption#1#2{%
  \vskip\abovecaptionskip
  \sbox\@tempboxa{{\captionfonts #1: #2}}%
 \ifdim \wd\@tempboxa >\hsize
    {\captionfonts #1: #2\par}
  \else
    \hbox to\hsize{\hfil\box\@tempboxa\hfil}%
  \fi
  \vskip\belowcaptionskip}
\makeatother   % Cancel the effect of \makeatletter

\usepackage{mciteplus}

\usepackage[colorlinks=true,      linkcolor=blue,      urlcolor=blue,      filecolor=blue,      citecolor=blue,
      pdfstartview=FitH,     pdfpagemode=UseNone,      bookmarksopen=true]{hyperref}  % LINKS
      
\usepackage[all]{hypcap}     %should be loaded AFTER hyperref, which should otherwise be loaded last.

\title{Two Twist Transition Amplitudes\\
Full Paper Collection}

\begin{document}

\numberwithin{equation}{section}

%%%%%%%%%%%%%%%%%%%%%%%%%%%%%%%%%%%%%%%%%%%%%%%%%%%%%%%%%%%%
%                       DEFINITIONS

\mathchardef\mhyphen="2D

%%%%%%%%%%%%%%%%%%%%%%%%%%%%%%%%%%%%%%%%%%%%%%%%%%%%%%%%%%%%
%                        Commands

%			Wick Contractions
\makeatletter
\newcommand{\contraction}[5][1ex]{%
  \mathchoice
    {\contraction@\displaystyle{#2}{#3}{#4}{#5}{#1}}%
    {\contraction@\textstyle{#2}{#3}{#4}{#5}{#1}}%
    {\contraction@\scriptstyle{#2}{#3}{#4}{#5}{#1}}%
    {\contraction@\scriptscriptstyle{#2}{#3}{#4}{#5}{#1}}}%
\newcommand{\contraction@}[6]{%
  \setbox0=\hbox{$#1#2$}%
  \setbox2=\hbox{$#1#3$}%
  \setbox4=\hbox{$#1#4$}%
  \setbox6=\hbox{$#1#5$}%
  \dimen0=\wd2%
  \advance\dimen0 by \wd6%
  \divide\dimen0 by 2%
  \advance\dimen0 by \wd4%
  \vbox{%
    \hbox to 0pt{%
      \kern \wd0%
      \kern 0.5\wd2%
      \contraction@@{\dimen0}{#6}%
      \hss}%
    \vskip 0.2ex%
    \vskip\ht2}}
\newcommand{\contracted}[5][1ex]{%
  \contraction[#1]{#2}{#3}{#4}{#5}\ensuremath{#2#3#4#5}}
\newcommand{\contraction@@}[3][0.06em]{%
  \hbox{%
    \vrule width #1 height 0pt depth #3%
    \vrule width #2 height 0pt depth #1%
    \vrule width #1 height 0pt depth #3%
    \relax}}
\makeatother

%			Environments
\newcommand{\be}{\begin{equation}} % Use eqnarray instead
\newcommand{\ee}{\end{equation}} % Use eqnarray insted
\newcommand{\bea}{\begin{eqnarray}\displaystyle}
\newcommand{\eea}{\end{eqnarray}}
\newcommand{\bt}{\begin{tabular}}
\newcommand{\et}{\end{tabular}}
\newcommand{\bs}{\begin{split}}
\newcommand{\es}{\end{split}}

%			Vacuum States (In Math Environments)
\newcommand{\nsnsket}{|0_{NS}\rangle^{(1)}\newotimes |0_{NS}\rangle^{(2)}}					% Double NS ket
\newcommand{\nsnsbra}{{}^{(1)}\langle 0_{NS}| \newotimes {}^{(2)}\langle 0_{NS}|}		% Double NS bra
\newcommand{\nsket}{|0_{NS}\rangle}																									% Single NS ket
\newcommand{\nsbra}{\langle 0_{NS}|}																								% Single NS bra
\newcommand{\nstket}{|0_{NS}\rangle_t}																										% t-plane NS ket
\newcommand{\nstbra}{{}_t\langle 0_{NS}|}																								% t-plane NS bra
\newcommand{\rmmket}{|0_R^-\rangle^{(1)}\newotimes |0_R^-\rangle^{(2)}}							% Double R- ket
\newcommand{\rmmbra}{{}^{(1)}\langle 0_{R,-}| \newotimes {}^{(2)}\langle 0_{R,-}|}	% Double R- bra
\newcommand{\rmket}{|0_R^-\rangle}																									% Single R- ket
\newcommand{\rmbra}{\langle 0_{R,-}|}																								% Single R- bra
\newcommand{\rmtket}{|0_R^-\rangle_t}																								% t-plane R- ket
\newcommand{\rmtbra}{{}_t\langle 0_{R,-}|}																					% t-plane R- bra
\newcommand{\rppket}{|0_R^+\rangle^{(1)}\newotimes |0_R^+\rangle^{(2)}}							% Double R+ ket
\newcommand{\rppbra}{{}^{(1)}\langle 0_{R,+}| \newotimes {}^{(2)}\langle 0_{R,+}|}	% Double R+ bra
\newcommand{\rpket}{|0_R^+\rangle}																									% Single R+ ket
\newcommand{\rpbra}{\langle 0_{R,+}|}																								% Single R+ bra
\newcommand{\rptket}{|0_R^+\rangle_t}																								% t-plane R+ ket
\newcommand{\rptbra}{{}_t\langle 0_{R,+}|}																					% t-plane R+ bra
\newcommand{\rpmket}{| 0_R^+\rangle^{(1')} \newotimes | 0_R^-\rangle^{(2')}}					% R+ on 1, R- on 2, ket
\newcommand{\rpmbra}{{}^{(1')}\langle 0_{R,+}| \newotimes {}^{(2')}\langle 0_{R,-}|}	% R+ on 1, R- on 2, bra
\newcommand{\rmpket}{| 0_R^-\rangle^{(1)} \newotimes | 0_R^+\rangle^{(2)}}					% R- on 1, R+ on 2, ket
\newcommand{\rmpbra}{{}^{(1)}\langle 0_{R,-}| \newotimes {}^{(2)}\langle 0_{R,+}|}	% R- on 1, R+ on 2, bra

% Commutator %
\newcommand\com[2]{[#1,\,#2]}

% Command Shortcuts - Vacuum States %
\newcommand{\nsutvket}{|0_{NS}\rangle^{(1)}\otimes |0_{NS}\rangle^{(2)}}
\newcommand{\nsutvbra}{{}^{(1)}\langle 0_{NS}| \otimes {}^{(2)}\langle 0_{NS}|}
\newcommand{\nstvket}{|0_{NS}\rangle}
\newcommand{\nstvbra}{\langle 0_{NS}|}
\newcommand{\nstpket}{|0\rangle_t}
\newcommand{\nstpbra}{{}_t\langle 0|}
\newcommand{\rmutvket}{|0_R^-\rangle^{(1)}\otimes |0_R^-\rangle^{(2)}}
\newcommand{\rmutvbra}{{}^{(1)}\langle 0_{R,-}| \otimes {}^{(2)}\langle 0_{R,-}|} 
\newcommand{\rmtvket}{|0_R^-\rangle}
\newcommand{\rmtvbra}{\langle 0_{R,-}|}
\newcommand{\rmtpket}{|0_R^-\rangle_t}
\newcommand{\rmtpbra}{{}_t\langle 0_{R,-}|}
\newcommand{\rputvket}{|0_R^+\rangle^{(1)}\otimes |0_R^+\rangle^{(2)}}
\newcommand{\rputvbra}{{}^{(1)}\langle 0_{R,+}| \otimes {}^{(2)}\langle 0_{R,+}|} 
\newcommand{\rptvket}{|0_R^+\rangle}
\newcommand{\rptvbra}{\langle 0_{R,+}|}
\newcommand{\rptpket}{|0_R^+\rangle_t}
\newcommand{\rptpbra}{{}_t\langle 0_{R,+}|}
\newcommand{\stp}{\sigma_2^+}
\newcommand{\stm}{\sigma_2^-}

% Greek Letters %
\renewcommand{\a}{\alpha}	% For spectral flow and boson modes
\renewcommand{\b}{\beta}
\newcommand{\g}{\gamma}		% For exponential coefficients
\newcommand{\G}{\Gamma}
\renewcommand{\d}{\delta}
\newcommand{\D}{\Delta}
\renewcommand{\c}{\chi}			% For final state
\newcommand{\C}{\Chi}
\newcommand{\p}{\psi}			% For fermion field
\renewcommand{\P}{\Psi}
\newcommand{\s}{\sigma}		% For twist operator and cylinder spatial coordinate
\renewcommand{\S}{\Sigma}
\renewcommand{\t}{\tau}		% For cylinder temporal coordinate
\newcommand{\e}{\epsilon}
\newcommand{\n}{\nu}
\newcommand{\m}{\mu}
\renewcommand{\r}{\rho}
\renewcommand{\l}{\lambda}
\newcommand{\sh}{\,\hat\sigma\,} % For sigma-hat with good spacing

%			Math
\newcommand{\nn}{\nonumber\\} 		% New line without numbering current line
\newcommand{\newotimes}{}  				% Change this to put tensor products back in.
\newcommand{\diff}{\,\text{d}}		% For differentials
\newcommand{\h}{{1\over2}}				% Shortcut for 1/2
\newcommand{\Gf}[1]{\G \Big{(} #1 \Big{)}}	% Gamma Function Shortcut
\newcommand{\floor}[1]{\left\lfloor #1 \right\rfloor}
\newcommand{\ceil}[1]{\left\lceil #1 \right\rceil}

%      Calligraphic Font
\def\cA{{\cal A}} \def\cB{{\cal B}} \def\cC{{\cal C}}
\def\cD{{\cal D}} \def\cE{{\cal E}} \def\cF{{\cal F}}
\def\cG{{\cal G}} \def\cH{{\cal H}} \def\cI{{\cal I}}
\def\cJ{{\cal J}} \def\cK{{\cal K}} \def\cL{{\cal L}}
\def\cM{{\cal M}} \def\cN{{\cal N}} \def\cO{{\cal O}}
\def\cP{{\cal P}} \def\cQ{{\cal Q}} \def\cR{{\cal R}}
\def\cS{{\cal S}} \def\cT{{\cal T}} \def\cU{{\cal U}}
\def\cV{{\cal V}} \def\cW{{\cal W}} \def\cX{{\cal X}}
\def\cY{{\cal Y}} \def\cZ{{\cal Z}}

%				Math Bold Face
\def\mC{\mathbb{C}} \def\mP{\mathbb{P}}  
\def\mR{\mathbb{R}} \def\mZ{\mathbb{Z}} 
\def\mT{\mathbb{T}} \def\mN{\mathbb{N}}
\def\mH{\mathbb{H}} \def\mX{\mathbb{X}}
\def\CP{\mathbb{CP}}
\def\RP{\mathbb{RP}}
\def\Z{\mathbb{Z}}
\def\N{\mathbb{N}}
\def\H{\mathbb{H}}

%			Quantum Mechanics Notation
\newcommand{\Zd}{\ensuremath{ Z^{\dagger}}}
\newcommand{\Xd}{\ensuremath{ X^{\dagger}}}
\newcommand{\Ad}{\ensuremath{ A^{\dagger}}}
\newcommand{\Bd}{\ensuremath{ B^{\dagger}}}
\newcommand{\Ud}{\ensuremath{ U^{\dagger}}}
\newcommand{\Td}{\ensuremath{ T^{\dagger}}}
\newcommand{\T}[3]{\ensuremath{ #1{}^{#2}_{\phantom{#2} \! #3}}}		%general tensor with upper indices first 
\newcommand{\tr}{\operatorname{tr}}
\newcommand{\sech}{\operatorname{sech}}
\newcommand{\Spin}{\operatorname{Spin}}
\newcommand{\Sym}{\operatorname{Sym}}
\newcommand{\Com}{\operatorname{Com}}
\def\adj{\textrm{adj}}
\def\id{\textrm{id}}
\def\pb{\ov\psi}
\def\pt{\widetilde{\psi}}
\def\at{\widetilde{\a}}
\def\cb{\ov\chi}
\def\db{\bar\partial}
\def\delb{\bar\partial}
\def\dbar{\ov\partial}
\def\dag{\dagger}
\def\dalpha{{\dot\alpha}}
\def\dbeta{{\dot\beta}}
\def\dgamma{{\dot\gamma}}
\def\ddelta{{\dot\delta}}
\def\ad{{\dot\alpha}}
\def\bd{{\dot\beta}}
\def\dg{{\dot\gamma}}
\def\dd{{\dot\delta}}
\def\th{\theta}
\def\Th{\Theta}
\def\eb{{\ov \epsilon}}
\def\gb{{\ov \gamma}}
\def\wb{{\ov w}}
\def\Wb{{\ov W}}
\def\D{\Delta}
\def\DD{\Delta^\dag}
\def\Db{\ov D}
\def\ov{\overline}
\def\Slash{\, / \! \! \! \!}
\def\dslash{\partial\!\!\!/} 
\def\Dslash{D\!\!\!\!/\,\,}
\def\fslash#1{\slash\!\!\!#1}
\def\Fslash#1{\slash\!\!\!\!#1}
\def\del{\partial}
\def\delb{\bar\partial}
\newcommand{\ex}[1]{{\rm e}^{#1}} 
\def\ii{{i}}
\newcommand{\vs}[1]{\vspace{#1 mm}}
\newcommand{\ve}{{\vec{\e}}}
\newcommand{\shalf}{\frac{1}{2}}
\newcommand{\lb}{\rangle}
\newcommand{\al}{\ensuremath{\alpha'}}
\newcommand{\ap}{\ensuremath{\alpha'}}
\newcommand{\ft}[2]{{\textstyle {\frac{#1}{#2}} }}

%				Other
\newcommand{\rmd}{\mathrm{d}}
\newcommand{\rmx}{\mathrm{x}}
\def\tA{ {\widetilde A} } 
\def\one{{\hbox{\kern+.5mm 1\kern-.8mm l}}}
\def\zero{{\hbox{0\kern-1.5mm 0}}}
\def\eq#1{(\ref{#1})}
\newcommand{\secn}[1]{Section~\ref{#1}}
\newcommand{\tbl}[1]{Table~\ref{#1}}
\newcommand{\fig}{Fig.~\ref}
\def\sqi{{1\over \sqrt{2}}}
\newcommand{\hsp}{\hspace{0.5cm}}
\def\half{{\textstyle{1\over2}}}
\let\ci=\cite \let\re=\ref
\let\se=\section \let\sse=\subsection \let\ssse=\subsubsection
\newcommand{\dpb}{D$p$-brane}
\newcommand{\dpbs}{D$p$-branes}
\def\gh{{\rm gh}}
\def\sgh{{\rm sgh}}
\def\NS{{\rm NS}}
\def\R{{\rm R}}
\def\Qp{Q_{\rm P}}
\def\QP{Q_{\rm P}}
\newcommand\dott[2]{#1 \! \cdot \! #2}
\def\eo{\overline{e}}
\newcommand{\bb}{\bigskip}
\newcommand{\ac}[2]{\ensuremath{\{ #1, #2 \}}}
\renewcommand{\ell}{l}
\newcommand{\z}{\ell}
\newcommand{\bm}{\bibitem}

\vspace{16mm}

 \begin{center}
{\LARGE One-Loop Transition Amplitudes\\ in the D1D5 CFT }

\vspace{18mm}
{\bf  Zaq Carson\footnote{carson.231@osu.edu}, Shaun Hampton\footnote{hampton.197@osu.edu}, and Samir D. Mathur\footnote{mathur.16@osu.edu}
\\}
\vspace{15mm}
Department of Physics,\\ The Ohio State University,\\ Columbus,
OH 43210, USA\\ 
\vspace{8mm}
\end{center}

\vspace{10mm}

\thispagestyle{empty}
\begin{abstract}

\vspace{3mm}

We consider the issue of thermalization in the D1D5 CFT. Thermalization is expected to correspond to the formation of a black hole in the dual gravity theory. We start from the orbifold point, where the theory is essentially free, and does not thermalize. In earlier work it was noted that there was no clear thermalization effect when the theory was deformed off the orbifold point to first order in the relevant twist perturbation. In this paper we consider the deformation to second order in the twist, where we do find effects that can cause thermalization of an initial perturbation. We consider a 1-loop process where two untwisted copies of the CFT are twisted to one copy and then again untwisted to two copies.  We start with a single oscillator excitation on the initial CFT, and compute the effect of the two twists on this state. We find simple approximate expressions for the Bogoliubov coefficients and the behavior of the single oscillator excitation in the continuum limit, where the mode numbers involved are taken to be much larger than unity.  We also prove a number of useful relationships valid for processes with an arbitrary number of twist insertions.
\end{abstract}
\newpage

\section{Introduction}\label{Intro}
Black holes are systems in which gravity serves as the dominant force, yet their evaporation is fundamentally quantum.  This provides us with a compelling testing ground for any attempt we make at understanding quantum gravity.  In the case of string theory, the gravitational description can be placed in a setting which affords us a CFT dual \cite{adscft}.  This dual CFT is the focus of our investigations.

While the exact dual CFT is strongly coupled, an examination of its 'free' or 'orbifold' point has garnered many fruitful results \cite{sv,lm1, lm2,orbifold1,orbifold2,deformation,orbifold3}.  At this coupling the dual CFT consists of several symmetrized copies of a free CFT whose target space is a 1+1 dimensional sigma model.  This orbifold model has successfully reproduced the entropy and greybody factors of near-extremal black holes  \cite{radiation1}, but it cannot provide a description of their formation.  This is because the black hole formation is dual to a thermalization of the dual CFT, which does not in general occur for excitations in a free theory.

In light of this, it is fruitful to explore the deformation of the CFT away from its orbifold point.  This deformation is given by the operator \cite{acm1}:
\bea
\hat{O}_{\dot A\dot B}\left(w_0,\bar{w}_0\right)=\left[{1\over2\pi i}\int_{w_0}G^-_{\dot A}(w)\diff w\right]\left[{1\over2\pi i}\int_{\bar{w}_0}\bar{G}^-_{\dot B}(\bar w)\diff \bar w\right]\s_2^{++}\left(w_0,\bar{w}_0\right).\label{FullDeformationOperator}
\eea
The index notations are detailed in appendix \ref{ap:CFT-notation}.  This operator has two key components.  The first is the twist, $\s_2$, which joins two copies of the free CFT.  If these copies were built on circles of length $2\pi R$, the twist merges them into a single CFT on a circle of length $4\pi R$. The second component is the supercharge operator $G$, which is applied in both left-moving and right-moving sectors.

As detailed in \cite{acm1,acm2}, the supercharge contours can be removed from the twist by stretching them away until they act on the initial and final states of the process.  This allows us to separate out the action of the 'bare twist' $\s_2$.  An assessment of this process at first order has been completed \cite{chmt1}.  At this order one could not see  thermalization of the initial excitations on the CFT. 

\begin{figure}[tbh]
\begin{center}
\includegraphics[width=0.3\columnwidth]{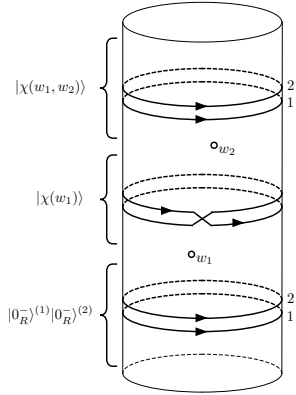}
\end{center}
\caption{The cylinder with twist insertions at $w_1$ and $w_2$.  Below the first twist we have the negative Ramond vacuum on each component string.  Above both twists we have the state $|\chi(w_1,w_2)\rangle$, which we will compute.  In the intermediate regions we have a single doubly-wound component string in the state $|\chi(w_1,w_2)\rangle$.  This state was computed in \cite{acm1} and is not used here.}
\label{figone}
\end{figure}

In \cite{chm1}, early results for a 1-loop computation were presented.  The particular second-order case under examination involved one twist joining two singly-wound copies of the orbifold CFT to a single copy living on a double circle, followed by a second twist which returns the double circle back to two singly-wound copies.  This process is shown in Figure \ref{figone}. It was shown that when the initial CFT copies are both in a vacuum state, the result is a squeezed state of the schematic form:
\bea
\s_2^+(w_2)\s_2^+(w_1)|0\rangle &=& e^{\g^B_{mn}\a_{-m}\a_{-n}+\g^F_{rs}d_{-r}d_{-s}}|0\rangle ~\equiv~ |\chi\rangle,
\eea
where the mode indices are summed over all creation operators.  The $\a$ modes are bosonic, while the $d$ modes are fermionic.  The coefficients $\g^B$ and $\g^F$ were expressed in terms of finite sums and their behavior for large indices was analyzed.

In the present paper, we extend the 1-loop analysis to the case where an initial excitation in present in the CFT state.  It is such initial excitations that are expected to change in the process of thermalization: if we start with an oscillator $\alpha_{-n}$, then a process of thermalization would convert this to other oscillator modes. By the general methods developed in earlier works on this subject, we expect that a single oscillator will, after the twist, go over to a linear combination of single oscillator states. This transition will be  characterized by a transition amplitude $f$:
\bea
\s_2^+(w_2)\s_2^+(w_1)\a_{-n}|0\rangle &=& \sum_{p}f^B_{np}\a_{-p}|\chi\rangle\nn
\s_2^+(w_2)\s_2^+(w_1)d_{-n}|0\rangle &=& \sum_{p}f^F_{np}d_{-p}|\chi\rangle,
\label{eqone}
\eea
where again our sum runs over all creation modes.  The full process of thermalization will then involve other pairs of oscillator excitations that arise from the exponential in oscillator pairs that arises from the transformation of the vacuum: if the transition (\ref{eqone}) takes the oscillator to a lower energy, then the remaining energy can be made up from such oscillator pairs from the exponential. Oscillators will also arise from the supercharges that have to be applied to the twist operator. We do not consider these supercharges in the present paper, since their action can be separated from the effect of the twists; we hope to return to an analysis of energy conservation and the complete created  state in a later work.

We note that the result (\ref{eqone}) is schematically similar to the first-order deformation studied in  \cite{acm2}.  Indeed, we expect this form at all orders for the same reason we expect the exponential form of $|\chi\rangle$ at all orders: Each mode on the cylinder maps to a linear combination of single modes with the same SU(2) indices in the twist-free covering space.  Since the vacuum gives an exponential of oscillator pairs, a  single initial excitation will give a single excitation above the exponential state arising from the vacuum.

Let us note the role of the computation of this paper in the more general program on thermalization that we are pursuing. We are looking for the essential vertex that leads to thermalization. We have found that such a vertex does not emerge at first order in perturbation theory, but we do expect it to emerge with the second order process we are now studying. In this paper we will compute an essential part of this vertex; i.e., the part where we have two twist deformations applied to a simple initial state. In a following paper, we will explain how to incorporate the supercharges; these will arise as contours applied to the initial and final states. Finally, one will have to integrate over the positions of the supercharges. But to understand the qualitative nature of the vertex can already be seen before this integration; i.e., once the effect of two deformation operators is known at arbitrary locations, we can see the distribution of energy modes that are produced by the thermalization vertex.

We note a couple of additional points. We are interested in the process of black hole formation. This process starts with a low entropy initial state, which we can just take to be a state with a single high energy particle falling in the throat of the D1D5 geometry. (We are considering the case where the spatial direction of the 1+1 dimensional CFT is compactified to a circle, as is the case for the D1D5P black hole; in this case a single high energy particle sent in towards $r=0$ will create a black hole in the classical computation.) This single particle can be described by one left and one right oscillator; this is the case we have taken in the present paper. In the limit where the oscillator mode number is high, we have a scaling of the kind studied in \cite{cmt}. Thus while we have taken singly wound strings in the present paper, the amplitude computation can be immediately scaled to the case where we have two strings with equal windings $N$ each, with $N$ arbitrarily large. This case would be of the kind that represents the long string states that are expected to dominate in the black hole phase of the D1D5 CFT. The only new features that come when going from singly wound strings to strings with winding $N$ is that with multiwound strings we have to allow the twist interaction to act on any of the $N$ strands. But this effect can be reproduced from the case with singly wound strings by going to the covering space of the multiply wound string. On this cover we take different spatial locations for the twist on the circle $0\le \sigma<2\pi$; i.e., we place images of the twist at points $\sigma= \sigma_0+{2\pi k/N}$ for $k=0, 1, \dots N-1$, and this takes into account interaction on different strands of the multiply wound string.  Thus a lot of information can be obtained by studying the two twist vertex with a single high energy oscillator on singly wound strings.

The plan of this paper is as follows.  In section \ref{CFT}, we introduce the orbifold CFT.  In section \ref{Relations}, we produce some general relationships for transition amplitudes at all orders of the twist operator.  One of these results relates the bosonic amplitude to the fermionic amplitude for nonzero mode indices.  In section \ref{tPlane}, we lay out the tools used in calculating the transition amplitudes for our specific 1-loop case.  In section \ref{Bosons} we calculate the bosonic amplitude and describe its behavior for large indices.  In section \ref{Fermions} we assess the case of fermion zero modes.  In section \ref{ContinuumLimit} we look at the continuum limit, where the mode numbers are much larger than unity.

\section{The orbifold CFT}\label{CFT}

Let us begin by recalling the orbifold CFT that we will be working with. Consider type IIB string theory, compactified as:
\bea
M_{9,1} &\to& M_{4,1}\times S^1\times T^4.
\eea
Now wrap $N_1$ D1 branes on $S^1$ and $N_5$ D5 branes on $S^1 \times T^4$.  We take $S^1$ to be large compared to $T^4$, so that the low energies are dominated by excitations only in the direction $S^1$.  This low-energy limit gives a $1+1$ dimensional CFT living on $S^1$.

At this point, variations in the moduli of string theory move us through the moduli space of the CFT on $S^1$.  It is conjectured that we can move to an 'orbifold point' where this CFT is a particularly simple sigma model \cite{orbifold2}.  We will begin in the Euclidean theory at this orbifold point.  The base space is a cylinder spanned by the coordinates $\t,\s$:
\bea
0\leq\s<2\pi,\qquad -\infty<\t<\infty.
\eea
The target space of this CFT is the symmetrized product of $N_1 N_5$ copies of $T^4$:
\bea
(T^4)^{N_1 N_5}/S_{N_1 N_5}.
\eea
Each copy gives 4 bosonic excitations and 4 fermionic excitations.  With an index $i$ ranging from 1 to 4, we label the bosonic excitations $X^i$, the left-moving fermionic excitations $\psi^i$, and the right-moving fermionic $\bar{\psi}^i$.  The total central charge is then $6 N_1 N_5$.

Fortunately, the twist operator fully factorizes into separate left-moving (holomorphic) and right-moving (antiholomorphic) sectors.  We thus constrain our analysis to the left-moving portion of (\ref{FullDeformationOperator}) and to holomorphic excitations.  The right-moving sector is completely analogous.

\subsection{NS and R vacuua}
At the orbifold point, each separate CFT copy has central charge $c=6$.  The lowest energy state of the left-moving sector for such a copy is the NS vacuum:
\bea
\nsket, &&\qquad h=0,\quad m=0,
\eea
where $h$ is the $L_0$ eigenvalue.  However, our interest lies mostly in the R sector of the CFT.  The vacuua of this sector are denoted by:
\bea
|0_R^{\pm}\rangle, &&\qquad h={1\over4}, \quad m = \pm \h\nn
|0_R\rangle,|\tilde 0_R\rangle, &&\qquad h={1\over4}, \quad m=0.\label{RamondVacuua}
\eea
These vacuua can be related through fermion zero modes, as detailed in appendix \ref{RVN}.  One can also relate the R and NS sectors via spectral flow \cite{spectralref}.  Spectral flow by a single unit in the left-moving sector produces the transformations:
\bea
\a=1: && \qquad \rmket \to \nsket, \quad \nsket \to \rpket\nn
\a=-1: && \qquad \rpket \to \nsket, \quad \nsket \to \rmket.
\eea
The other R vacuua can flow to the NS sector by first relating them to $|0_R^{\pm}\rangle$ via fermion zero modes.

\subsection{Copy Notation}
The full orbifold CFT contains a large number ($N_1 N_5$) of identical CFTs, each living on a single circle of circumference $2\pi R$.  We call these copies 'singly-wound.'  At arbitrary orders of the deformation (\ref{FullDeformationOperator}) the copies can be joined in any combination, resulting in CFTs whose compact dimension is any integer multiple of $2\pi R$.  However, when a twist $\s_2$ is applied to two windings of the same multi-wound CFT, that CFT splits into two separate CFTs with smaller winding number.  This allows us to form 1-loop processes in which some combination of CFTs are twisted together before returning back to their original winding configurations.

Most of this paper will address the simplest 1-loop case where two singly-wound CFTs are joined and then split.  However, we also wish to prove some transition amplitude relations for all orders of $\s_2$.  It is thus conducive to introduce some notation for handling the copy-identification indices in our computations.

From here on out, we will label an arbitrary CFT copy with the index $(k)$.  If this CFT is specifically before the twist operators of interest, we'll use the index $(i)$ instead.  The index $(j)$ will be used if the CFT is specifically after the twisting. In short:
\bea
(i) \implies \text{initial copy}, \quad (j) \implies \text{final copy}, \quad (k) \implies \text{any copy}.
\eea
If we need to specify the copy number explicitly, we will use a prime for copies located after the twists.  So $(1)$ means Copy 1 before the twists, while $(1')$ means Copy 1 after the twists.  We will never need to talk about copies in-between twist operators of interest.

Each copy $(k)$ can have its own winding number, denoted as $N_{(k)}$.  By the nature of the twist interactions, total winding number is conserved.  We express this as:
\bea
\sum_{(j)}N_{(j)} &=& \sum_{(i)}N_{(i)}.
\eea

We will henceforth use the symbol $\hat\sigma$ to denote whatever arbitrary combinations of pairwise twists we wish to apply.  The initial vacuum will be a tensor product of Ramond vacuua across all initial copies.  In general, we will denote such a vacuum (before and after the twist) by:
\bea
|\emptyset\rangle &\equiv& \prod_{(i)}|0^*_R\rangle^{(i)}\nn
|\emptyset'\rangle &\equiv& \prod_{(j)}|0^*_R\rangle^{(j)},
\eea
where the notation $|0^*_R\rangle$ is used to indicate an unspecified type of R vacuum.  The four possibilities were detailed in Equation (\ref{RamondVacuua}).  In general, the type of R vacuum need not be the same between the different copies.  The twisted vacuum is then:
\bea
\hat\s\,|\emptyset\rangle &\equiv& |\chi\rangle.
\eea

Our bosonic and fermionic fields are defined in a manner consistent with \cite{acm1, acm2}.  This gives the following (anti)commutation relations:
\bea
\left[\a_{A\dot A,m}^{(k)},\a_{B\dot B,n}^{(k')}\right] &=& -n\e_{AB}\e_{\dot A\dot B}\d^{(k)(k')}\nn
\left\{d_{m}^{(k),\a A},d_{n}^{(k'),\b B}\right\} &=& -N_{(k)}\e^{\a\b}\e^{AB}\d^{(k)(k')}.\label{CommutationRelations}
\eea
In our notation, the mode numbers $m, n$ are integers.
We also have a general supercurrent that can be used to form supercharge modes.  For each copy (k), the modes are defined in terms of a supercurrent contour.  However, we can also write the supercharge modes in terms of bosonic and fermionic modes:
\bea
G^{(k),\a}_{\dot A,n} &=& -{i\over N_{(k)}}\sum_m d^{(k),\a A}_m\a_{A\dot A,n-m}^{(k)}.
\eea
We will also make use of the `full $G$' operator, which is defined such that:
\bea
G^{\a}_{\dot A,n} &=& \begin{cases}
\sum\limits_{(i)}G^{(i),\a}_{\dot A,n} & \text{before twists}\\
\sum\limits_{(j)}G^{(j),\a}_{\dot A,n} & \text{after twists}.
\end{cases}
\eea
While this operator has been stripped away from the twist operators that we consider in this paper, the supercurrent is still useful for proving certain transition amplitude relationships in full generality.

Lastly, we define the transition amplitudes $f$ as follows:
\bea
\hat \s \, \a^{(i)}_{A\dot A,-n}|\emptyset\rangle &=& \sum_{(j)}\sum_{p} f^{B,(i)(j)}_{np}\a^{(j)}_{A\dot A,-p}|\chi\rangle\nn
\hat \s \, d^{(i),\a A}_{-n}|\emptyset\rangle &=& \sum_{(j)}\sum_{p} f^{F\a,(i)(j)}_{np}d^{(j),\a A}_{-p}|\chi\rangle.
\eea
The mode number $p$ also runs over integers.
Here the $(j)$ sum spans all final copies while the $p$ sum spans all values for which the corresponding operator does not annihilate the particular R vacuum upon which $|\chi\rangle$ is built.  We anticipate two distinct cases for the fermions since their SU(2) R charge $\a$ is the same type of charge carried by the twist operators.  All other group indices should be symmetric under the twists.

\section{Relations between transition amplitudes}\label{Relations}

Our interest is in computing the following kind of amplitude. We start with either the vacuum state or a state containing some oscillator excitations. We then apply a certain number of twist operators $\hat\sigma$ at definite locations. We then ask for the final state generated by this procedure. The actual deformation operator taking the CFT away from the orbifold point also contains contours of the supercharge $G$, but as mentioned before, these contours can be pulled away to act on the initial and final states, so a principal nontrivial part of the computation involves the effect of the twists $\hat \sigma$.

We have already mentioned some general properties of the states generated by the twists. Acting on the vacuum $|0\rangle$, the action of any number of twists is given by the form $\exp[\gamma^B\alpha\alpha+\gamma^F dd] |0\rangle$, where the $\alpha$ and $d$ are bosonoc and fermionic oscillatirs respectively, and $\gamma^B, \gamma^F$ are Bogoliubov coefficients that we need to compute. If we had a single oscilaltor in the intial state, then we get the exponential as before, but the oscillator is changed to a linear combination of operators (\ref{eqone}). If there is more than one oscillator in the initial state, then one gets the same behavior for individual oscillators, but in addition one finds Wick contractions between pairs of operators present in the initial state; examples of this were computed in \cite{acm2}.

In this section, we will derive some general relationships governing the action of twist operators. These relations reduce the effort involved in computing the relevant amplitudes, as they relate some amplitudes to others.  

\subsection{Transpose Relation for Bosons}\label{TransposeRelations}
Consider a situation in which the winding configuration of the final state are identical to that of the initial state.  That is, we have the relation:
\bea
(i) = (j) &\implies& N_{(i)} = N_{(j)},
\eea
for all $(i),(j)$.  Now consider the amplitude:
\bea
\mathcal{A} &\equiv& \langle\emptyset'| \a^{(j)}_{++,n}\sh\a^{(i)}_{--,-m}|\emptyset\rangle.
\eea
Passing the initial boson through the twists, we find:
\bea
\mathcal{A} &=& \sum_{(j'),p}f^{B,(i)(j')}_{mp}\langle\emptyset'| \a^{(j)}_{++,n}\a^{(j')}_{--,-p}\sh|\emptyset\rangle.
\eea
We now apply the commutation relations (\ref{CommutationRelations}) to obtain:
\bea
\mathcal{A} &=& -n f^{B,(i)(j)}_{mn}\langle\emptyset'|\sh|\emptyset\rangle.\label{Step1}
\eea

Now the bosonic operators have no SU(2) R charge.  This means their behavior is in general independent of both the choice of particular R vacuua as well a the charge of the twists. The first independence allows us to choose any combination of R vacuua for both $|\emptyset\rangle$ and $\langle\emptyset'|$ without altering the amplitude $\mathcal{A}$.  Since both cover the same winding configurations, one possible choice is to swap the two vacuum choices.  We thus find:
\bea
\mathcal{A} &=& \sum_{(j'),p}f^{B,(i)(j')}_{mp}\langle\emptyset| \a^{(j)}_{++,n}\sh\a^{(j')}_{--,-p}|\emptyset'\rangle.\label{ReversedAmp}
\eea

We now make use of the independence of the twist operator's SU(2) R charge.  This means that for bosons $\hat\s^{\dagger}=\hat\s^{-1}$, the reverse twisting process.\footnote{In general one would need also apply a global interchange of SU(2) R charges for this relation.}  We thus take the conjugate of both sides in Equation (\ref{ReversedAmp}) to obtain:
\bea
\mathcal{A}^* &=& \langle\emptyset'| \a^{(i)}_{++,m}\,\hat{\s}^{-1}\,\a^{(j)}_{++,-n}|\emptyset\rangle\nn
&=& -m\left( \tilde{f}^{B,(j)(i)}_{nm}\right)^*\langle\emptyset'|\sh|\emptyset'\rangle,\label{GeneralTransposeRelation}
\eea
where here the tilde denotes the transition amplitude for the reversed process.  Comparing to Equation (\ref{Step1}), we find:
\bea
n f^{B,(i)(j)}_{mn}&=& m \left(\tilde f^{B,(j)(i)}_{nm}\right)^*
\eea

For the particular one-loop case that we examine later, the twisting process is its own reversal (for bosons) and the transition amplitudes are real-valued (for convenient Minkowski coordinates). Furthermore, swapping the copy indices on $f$ amounts to either a global copy redefinition or nothing, neither of which has any physical effect.  Thus (\ref{GeneralTransposeRelation}) becomes:
\bea
nf^{B,(i)(j)}_{mn} &=& mf^{B,(i)(j)}_{nm}.\label{TwoTwistTransposeRelation}
\eea

\subsection{Supersymmetry Relations}
Here we find relations between the bosonic and fermionic transition amplitudes for arbitrary twisting.  To show these relations, we will make use of the fact that a $G^+$ current wrapped around any $\s_2^+$ operator with no extra weight in the integrand vanishes:
\bea
\oint_{w_0} G^{+}_{\dot A}(w)\,\s_2^+(w_0)\diff w &=& 0.
\eea
This can be seen by mapping to an appropriate covering plane, as shown in appendix \ref{GPlusProof}.  At the same time, one can deform the contour to obtain:
\bea
\oint_{w_0} G^{+}_{\dot A}(w)\,\s_2^+(w_0)\diff w &=& G^+_{\dot A,0}\,\s_2^+(w_0) -\s_2^+(w_0)G^+_{\dot A,0}\nn
&=& \left[G^+_{\dot A,0},\s_2^+(w_0)\right]~=~0.
\eea
Thus the full $G^+_{\dot A,0}$ mode commutes with the basic twist operator $\s_2^+$.  Since any twist operation of interest can be constructed from a combination of $\s_2^+$ operators, we find:
\bea
G^+_{\dot A,0}\,\hat\s &=& \hat\s\, G^+_{\dot A,0}.
\eea

Now for each copy, that copy's $G^+$ zero mode annihilates  the Ramond vacuua; thus in particular it annihilates the negative Ramond vacuum.  We thus have:
\bea
G^+_{\dot A,0}|\chi\rangle &=& G^+_{\dot A,0}\sh|\emptyset\rangle\nn
&=& \hat\s\,G^+_{\dot A,0}|\emptyset\rangle\nn
&=& 0.
\eea

We will now apply these relations to the two cases of possible SU(2) R charges for fermion modes.

\subsubsection{Fermions with negative SU(2) R charge}
Let us here consider the state with an initial negative-charge fermion mode and act on it with a $G^+$ zero mode.  Since this zero mode commutes with all twists, we have for instance:
\bea
G^+_{+,0}\sh d^{(i),--}_{-n}|\emptyset\rangle &=& \hat\s\,G^+_{+,0}d^{(i),--}_{-n}|\emptyset\rangle.\label{MinusRelationStart}
\eea

For $n=0$ the relationship is trivial: both sides vanish.  Setting aside such zero modes, we proceed with (\ref{MinusRelationStart}) in two ways. Starting with the left-hand side, we find:
\bea
G^+_{+,0}\sh d^{(i),--}_{-n}|\emptyset\rangle &=&\sum_{p,(k)} G^+_{+,0}f^{F-,(i)(j)}_{np}d^{(j),--}_{-p}|\chi\rangle\nn
&=& \sum_{p,(j)}f^{F-,(i)(j)}_{np}\left\{G^+_{+,0},d^{(j),--}_{-p}\right\}|\chi\rangle\nn
&=& \sum_{p',p,(j),(j'),A}\!\!\!\!\!\!\!f^{F-,(i)(j)}_{np}\left({-i\over N_{(j')}}\right)\left\{d^{(j'),+A}_{p'},d^{(j),--}_p\right\}\a_{A+,-p'}^{(j')}|\chi\rangle\nn
&=& -i\sum_{p,(j)}{1\over N_{(j)}}f^{F-,(i)(j)}_{np}\left(-N_{(j)}\right)\e^{+-}\e^{+-}\a^{(j)}_{++,-p}|\chi\rangle\nn
&=& i\sum_{p,(j)}f^{F-,(i)(j)}_{np}\a^{(j)}_{++,-p}|\chi\rangle.\label{NegativeLeftPart}
\eea

Now we turn to the right-hand side of (\ref{MinusRelationStart}) and perform a similar manipulation:
\bea
\s^+G^+_{+,0}d^{(i),--}_{-n}|\emptyset\rangle &=& \hat\s\,\left\{G^+_{+,0},d^{(i),--}_{-n}\right\}|\emptyset\rangle\nn
&=& \hat\s\,\sum_{(i'),n'}\left({-i\over N_{(i')}}\right)\left\{d^{(i'),+A}_{n'},d^{(i),--}_{-n}\right\}\a_{A+,-n'}^{(i')}|\emptyset\rangle\nn
&=& -{i\over N_{(i)}}\s^+\left(-N_{(i)}\right)\e^{+-}\e^{+-}\a_{++,-n}^{(i)}|\emptyset\rangle\nn
&=& i\sh\a_{++,-n}|\emptyset\rangle\nn
&=& i\sum_{p,(j)}f^{B,(i)(j)}_{np}\a_{++,-p}^{(j)}|\chi\rangle.\label{NegativeRightPart}
\eea
Combining this with (\ref{NegativeLeftPart}) we find that for any arbitrary twisting:
\bea
f^{B,(i)(j)}_{np} &=& f^{F-,(i)(j)}_{np},\qquad n,p>0.\label{NegativeRelation}
\eea

\subsubsection{Fermion with positive SU(2) R charge}
Let us now consider the state with an initial boson mode and act on it with a $G^+_0$ mode.  Since this zero mode commutes with all twists, we have for instance:
\bea
G^+_{+,0}\sh\a_{--,-n}^{(i)}|\emptyset\rangle &=& \hat\s\,G^+_{+,0}\a_{--,-n}^{(i)}|\emptyset\rangle.\label{PositiveRelationStart}
\eea

Again we set aside the $n=0$ case since it vanishes trivially.  We proceed with (\ref{PositiveRelationStart}) in the same two ways. Starting with the left-hand side, we find:
\bea
G^+_{+,0}\sh\a_{--,-n}^{(i)}|\emptyset\rangle &=& \sum_{p,(j)} G^+_{+,0}f^{B,(i)(j)}_{np}\a_{--,-p}^{(j)}|\chi\rangle\nn
&=& \sum_{p,(k)}f^{B,(i)(j)}_{np}\left[G^+_{+,0},\a_{--,-p}^{(j)}\right]|\chi\rangle\nn
&=& \sum_{p',p,(i),(j'),A}f^{B,(i)(j)}_{np}\left({-i\over N_{(j')}}\right)\left[\a_{A+,p'}^{(j')},\a_{--,-p}^{(j)}\right]d^{(j'),+A}_{-p'}|\chi\rangle\nn
&=& i\sum_{p,(j)}\left({-i\over N_{(j)}}\right)f^{B,(i)(j)}_{np}\left(-p\right)\e_{+-}\e_{+-}d^{(j),++}_{-p}|\chi\rangle\nn
&=& i\sum_{p,(j)}{p\over N_{(j)}}f^{B,(i)(j)}_{np}d^{(j),++}_{-p}|\chi\rangle.\label{PositiveLeftPart} 
\eea

Now we turn to the right-hand side of (\ref{PositiveRelationStart}) and perform a similar manipulation:
\bea
\hat\s\,G^+_{+,0}\a_{--,-n}^{(i)}|\emptyset\rangle &=& \hat{\s}\left[G^+_{+,0},\a_{--,-n}^{(i)}\right]|\emptyset\rangle\nn
&=& \hat\s\,\sum_{(i'),n'}\left({-i\over N_{(i')}}\right)\left[\a_{A+,n'}^{(i')},\a_{--,-n}^{(i)}\right]d^{(i'),+A}_{-n'}|\emptyset\rangle\nn
&=& -{i\over N_{(i)}}\sh\left(-n\right)\e_{+-}\e_{+-}d^{(i),++}_{-n}|\emptyset\rangle\nn
&=& i{n\over N_{(i)}}\sh d^{(i),++}_{-n}|\emptyset\rangle\nn
&=& i{n\over N_{(i)}}\sum_{p,(j)}f^{F+,(i)(j)}_{np}d^{(j),++}_{-n}|\chi\rangle.\label{PositiveRightPart}
\eea
Combining this with (\ref{PositiveLeftPart}) we find that for any arbitrary twisting:
\bea
{p\over N_{(j)}}f^{B,(i)(j)}_{np} &=& {n\over N_{(i)}}f^{F+,(i)(j)}_{np}, \qquad n,p>0.\label{PositiveRelation}
\eea

\section{The 1-Loop Case in Question}\label{tPlane}

Let us now come to the amplitude that we will compute in detail. We consider a 1-loop process that involves two twist operators; the first operator joins two singly wound copies to a doubly wound copy, and the second twist brings us back to two
 singly wound copies.

Thus the twist operator in question is described by the twists:
\bea
\hat{\s} &=& \s_2^+(w_2)\s_2^+(w_1),
\eea
for which we denote:
\bea
|\chi\rangle &=& \s_2^+(w_2)\s_2^+(w_1)|\emptyset\rangle\nn
|\emptyset\rangle &=& \rmmket.
\eea
This process is depicted in Figure \ref{figone}.  The state $|\chi\rangle$ was calculated in \cite{chm1}.

We consider a single oscillator excitation present in the initial state.  As argued previously, the final state will take the schematic form:
\bea
\hat \s \, \a^{(i)}_{A\dot A,-n}|\emptyset\rangle &=& \sum_{(j)}\sum_{p} f^{B,(i)(j)}_{np}\a^{(j)}_{A\dot A,-p}|\chi\rangle\nn
\hat \s \, d^{(i),\a A}_{-n}|\emptyset\rangle &=& \sum_{(j)}\sum_{p} f^{F\a,(i)(j)}_{np}d^{(j),\a A}_{-p}|\chi\rangle,\label{GeneralForm}
\eea
where we have now included all relevant indices.  The copy indices $(i)$ and $(j)$ can each take a value of 1 or 2 (we will apply primes for final state copies), while the charge index $\a$ can be $\pm \tfrac{1}{2}$.  In principal this gives twelve transition amplitudes, but we will use the relationships found in Section \ref{Relations} along with some additional symmetries to greatly reduce this proliferation.  

In calculating each of our transition amplitdes $f$, we use the same mathematical techniques presented in \cite{acm2}.  The cylinder shown in Figure \ref{figone} is parametrized by:
\bea
w = \t + i\s,
\eea
where $\t$ is a Euclideanized time coordinate and $\s$ is a compact spatial coordinate.  A factor of the cylinder's radius has been incorporated into each coordinate so that $\s$ is simply an angle.  The region $\t_1 < \t < \t_2$ has the two initial copies joined together to form a doubly wound CFT. We do not analyze this region in any detail.  Outside this region we find two singly wound CFTs.

We now map this configuration to the complex plain with coordinate $z$, and then map the complex plane to a double cover of itself with coordinate $t$.  This effectively maps the cylinder to a double cover of itself, allowing our bosonic fields to be single-valued in the $t$ plane.  The cover then has no twist insertions and we can proceed with the analysis in a straightforward fashion.

While coordinate shifts alone suffice for handling the bosons, we will need to address the anti-periodicity of the fermions.  The initial state has two singly-wound CFTs in the R sector.  These can be written in terms of spin fields:
\bea
\rmmket &=& S^{(1)-}(\t = -\infty)S^{(2)-}(\t = -\infty)\nsnsket.
\eea
Each twist operator also contains a spin field:
\bea
\s_2^+(w) &=& S^+(w)\s_2(w).
\eea
We will also find spin fields in the states we choose to cap with when calculating amplitudes.  These spin fields will occur in the out states, at $\t = \infty$.  All of these spin fields will be removed via spectral flow.

After mapping into the $t$ plane and removing all spin fields via spectral flows, we will be left with the local NS vacuum at both twist insertion locations.  This renders all of our fields single-valued at these points.  Our in and out states will map to particular neighborhoods within the $t$ plane.  The initial excitation forming a contour around the in state of the appropriate copy, while we will have a local NS vacuum at the other in state.  We can then deform the initial mode's contour smoothly through both twist locations as well as the other in state.  This allows us to bring the initial excitation to our out states in the $t$ plane.  We then invert our spectral flows and coordinate maps in order to move back to the cylinder.

The remainder of this section is divided into four parts.  First we outline the coordinate changes used to map the cylinder into a double cover of the complex plane and identify the images of all critical points.  Next we introduce the bosonic and fermionic mode operators, both for modes on the cylinder and for those natural to the $t$ plane.  Thirdly, we present the capping states motivated by the general form found in equation (\ref{GeneralForm}).  Finally, we assess the images of all spin fields to determine which spectral flows we need.

\begin{figure}[tbh]
\begin{center}
\includegraphics[width=0.4\columnwidth]{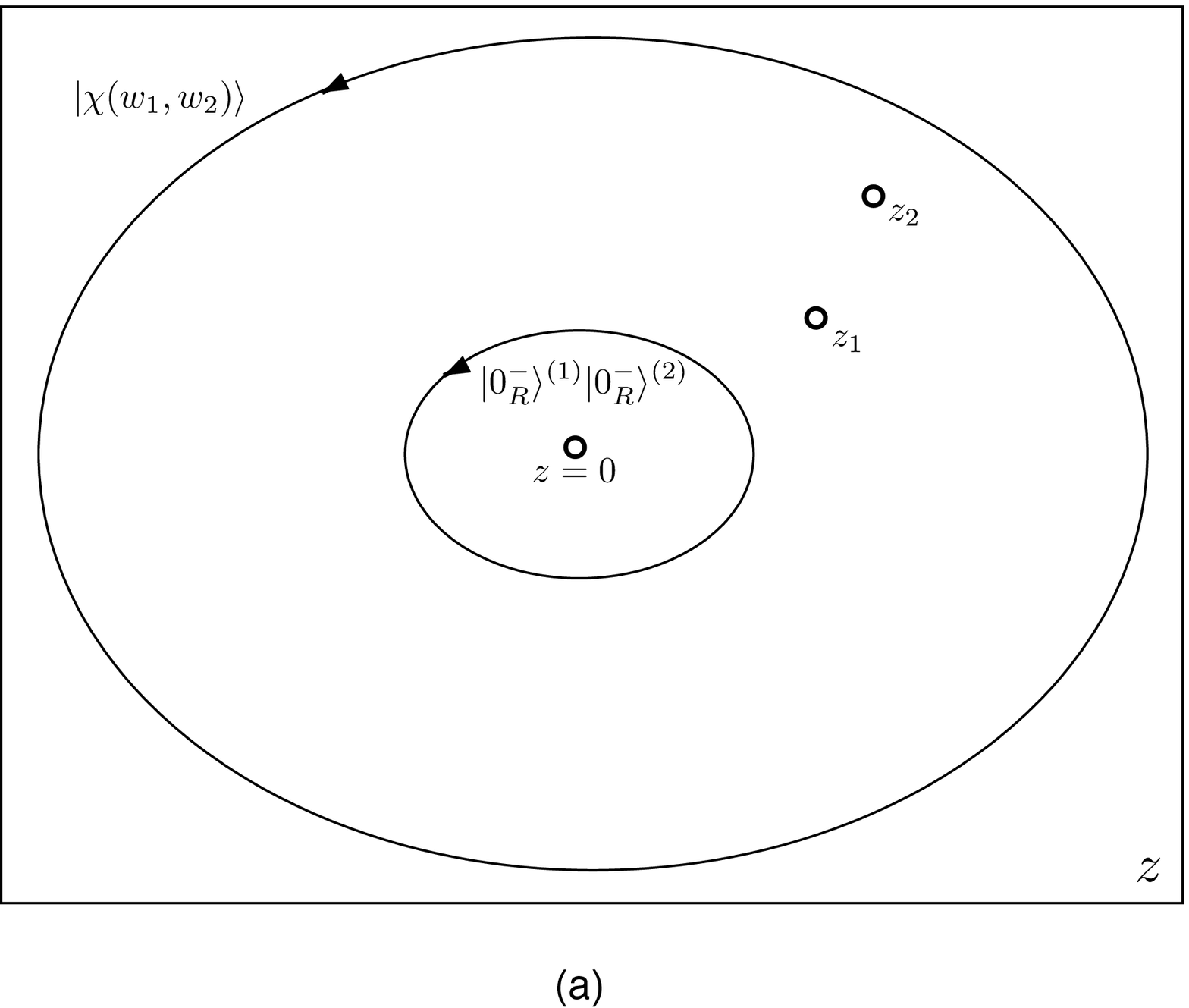} $\qquad\qquad$ \includegraphics[width=0.4\columnwidth]{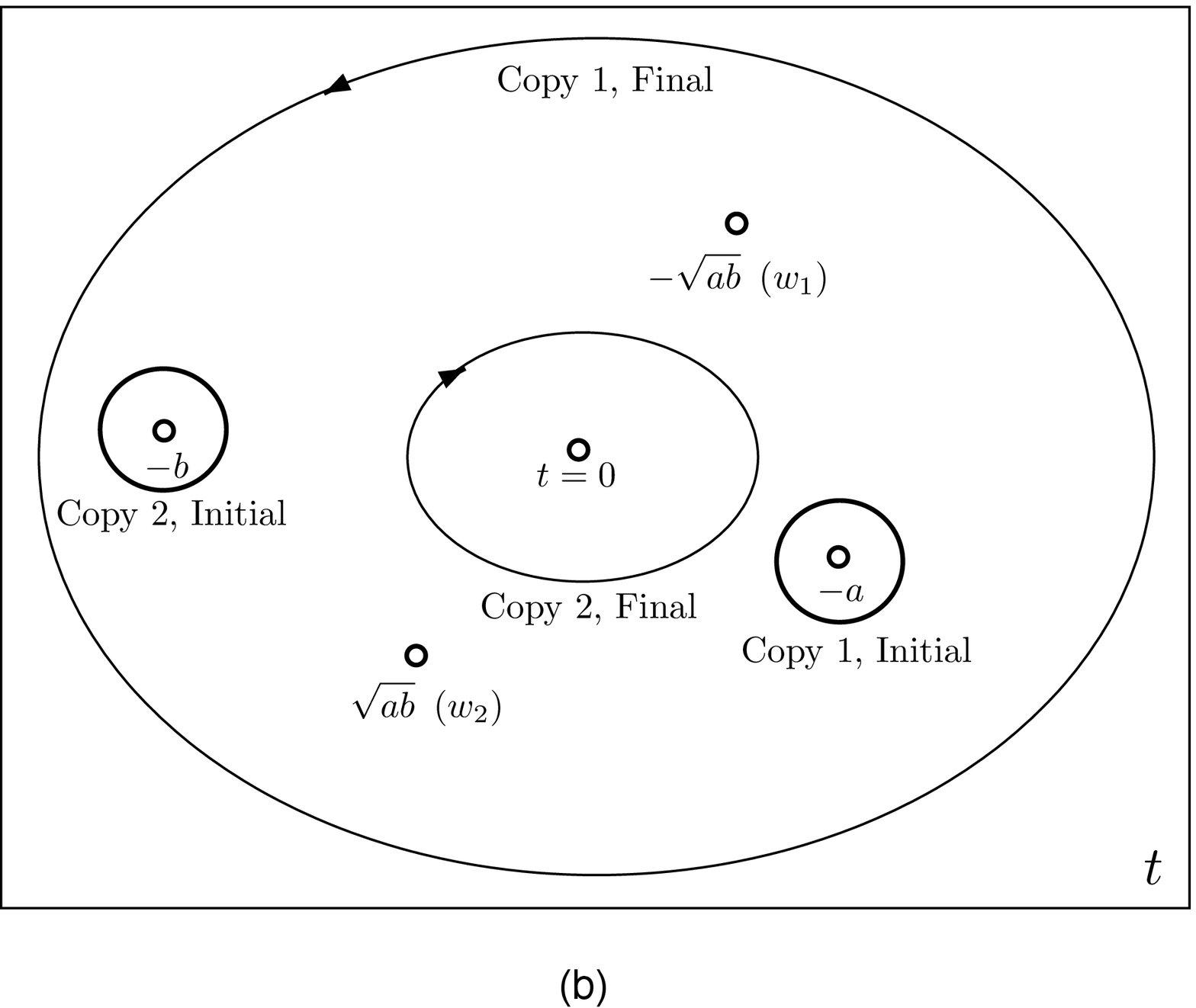}
\end{center}
\caption{The $z$ plane (a) and $t$ plane (b) with all the relevant image points labeled.  The intermediate state $|\chi(w_1)\rangle$ is not depicted.  In the $z$ plane the $\tau$ coordinate maps to the radial coordinate, while the $\s$ coordinate maps to the phase.  In the $t$ plane there are no simple directions corresponding to $\t$ and $\s$.}
\label{CoordinateMapFigure}
\end{figure}

\subsection{Coordinate Maps}
The coordinate maps are illustrated in Figure \ref{CoordinateMapFigure}.  The first map takes the cylinder to the complex plane via:
\bea
z &=& e^w ~=~ e^{\t+i\s}.\label{zMap}
\eea
Here the timelike $\tau$ because a radial component while the spacelike $\s$ becomes a phase.  The in states at $\tau = -\infty$ map to the origin of the $z$ plane while the out states at $\tau = \infty$ map to $z=\infty$.  The fields are still double-valued in these regions.  Since $\t_2 > \t_1$, we have:
\bea
|z_2| &=& e^{\t_2} ~>~ |z_1| ~=~ e^{\t_1}.
\eea

We now map into a double cover of the complex plane in a manner that separates out the two distinct CFTs occuring before and after the twists.  We use the same map presented in \cite{chm1}:
\bea
z &=& {(t+a)(t+b)\over t}.\label{tMap}
\eea
Here we have split the locations of the two copies in the in and out states.  Using primes for the copy indices of the out states, we have:
\bea
(1) &\to& z=0 ~\to~ t = -a \qquad z \sim t+a\nn
(2) &\to& z=0 ~\to~ t = -b \qquad z \sim t+b\nn
(1') &\to& z=\infty ~\to~ t = \infty \qquad z \sim t\nn
(2') &\to& z=\infty ~\to~ t = 0 \qquad z \sim t^{-1}.\label{ImagePoints}
\eea
We have made the choice of $-a$ as the image for copy 1 of the in state in order to remain consistent with  \cite{chm1}.

Since the twist operators carry spin fields, we will need to know their image points.  These are the bifurcation points of the double cover:
\bea
{\diff z\over \diff t} &=& 1 - {ab\over t^2} ~=~ 0\nn\nn
t_1 &=& -\sqrt{ab}\nn
t_2 &=& \sqrt{ab},
\eea
where we have again remained consistent with the choices in  \cite{chm1} for the choice of branch in $\sqrt{ab}$.  The full behavior of the twist insertion points is then:
\bea
e^{w_1} &=& z_1 ~=~ a+b-2\sqrt{ab}\nn
e^{w_2} &=& z_2 ~=~ a+b+2\sqrt{ab}.\label{BifurcationPoints}
\eea

\subsection{Mode operators}
We first introduce the bosonic and fermionic modes that live on the cylinder.  We work only with modes that appear for the regions $\t<\t_1$ and $\t>\t_2$ in the R sector.  Here both boson and fermion modes have integer mode number $n$.
\bea
\a^{(i)}_{A\dot A,n} &=& {1\over 2\pi}\int_{\t<\t_1,\s=0}^{\s=2\pi} \partial X^{(i)}_{A\dot A}(w)e^{nw}\diff w\nn
\a^{(j)}_{A\dot A,n} &=& {1\over 2\pi}\int_{\t>\t_2,\s=0}^{\s=2\pi} \partial X^{(j)}_{A\dot A}(w)e^{nw}\diff w\nn
d^{(i),\a A}_n &=& {1\over 2\pi i}\int_{\t<\t_1,\s=0}^{\s=2\pi} \psi^{(i),\a A}(w)e^{nw}\diff w\nn
d^{(j),\a A}_n &=& {1\over 2\pi i}\int_{\t>\t_2,\s=0}^{\s=2\pi} \psi^{(j),\a A}(w)e^{nw}\diff w.
\eea

We never have occasion to assess any commutation relations between initial and final modes as the two are inherently at \emph{unequal} times.  For modes that both occur on the same side of the twist operators, the commutation relations are:
\bea
[\alpha_{A\dot A, n}^{(k_1)f}, \alpha_{B\dot B, m}^{(k_2)f}] &=& -n\epsilon_{AB}\epsilon_{\dot A\dot B}\, \delta^{(k_1)(k_2)} \delta_{n+m,0}\nn
\left \{ d^{(k_1)f,\a A}_{r}, d^{(k_2)f,\b B}_{s} \right \} &=& -\e^{\a\b}\e^{AB}\d^{(k_1)(k_2)}\d_{r+s,0}\,.
\eea

We should also note a subtlety one encounters when mapping these cylinder modes into the $t$ plane.  Since we have chosen the final copy 1 to map to large $t$, copy 1 modes will always come to the left of copy 2 modes in the $t$ plane.  We will thus write all of our modes in this order even when they live on the cylinder.  This is an important convention for maintaining consistency when working with fermions.

We will also need to work with several different modes natural to the $t$ plane and built upon the NS vacuum.  In fact we need modes natural to the regions neighboring the image points of our in and out states on both copies.  For modes natural to the neighborhood of a point $t_0$, we write:
\bea
\tilde{\a}^{t\to t_0}_{A\dot A,n} &=& {1\over 2\pi} \oint\limits_{t=t_0} \partial_t X_{A\dot A}(t) t^n \diff t \label{BosontMode}\\
\tilde{d}^{\a A,t\to t_0}_r &=& {1\over 2\pi i} \oint\limits_{t=t_0} \psi^{\a A}(t) t^{r-\h} \diff t. \label{FermiontMode}
\eea
Here the fermion index is a half-integer, since we are in the NS sector.  The commutation relations are only simple when the modes are natural to the same neighborhoods.  We then have:
\bea
\left[\tilde{\a}^{t\to t_0}_{A\dot{A},m},\tilde{\a}^{t\to t_0}_{A\dot{A},n}\right]=-\e_{AB}\e_{\dot{A}\dot{B}}m\delta_{n+m,0} \label{bosoncommutation}\\
\left\{\tilde{d}^{\a A,t\to t_0}_q,\tilde{d}^{\a A,t\to t_0}_r\right\}=-\e^{\a\b}\e^{AB}\delta_{q+r,0}\;. \label{fermioncommutation}
\eea
We also have the natural behavior:
\bea
\tilde\a^{t\to t_0}_{A\dot A,n}\nstket &=& 0, \qquad n\geq0\nn
\tilde{d}^{\a A,t\to t_0}_q\nstket &=& 0, \qquad q>0.
\eea

\subsection{The Capping States}
Let us first re-present Equation \ref{GeneralForm}:
\bea
\hat \s \, \a^{(i)}_{A\dot A,-n}|\emptyset\rangle &=& \sum_{(j)}\sum_{p} f^{B,(i)(j)}_{np}\a^{(j)}_{A\dot A,-p}|\chi\rangle\nn
\hat \s \, d^{(i),\a A}_{-n}|\emptyset\rangle &=& \sum_{(j)}\sum_{p} f^{F\a,(i)(j)}_{np}d^{(j),\a A}_{-p}|\chi\rangle.\label{GeneralForm2}
\eea
In  \cite{chm1} it was found that the physical part of $|\chi\rangle$ can be written as:
\bea
|\chi\rangle &=& C(w_1,w_2)\,e^{\hat{Q}(w_1,w_2)}\,\rpmket\nn
&=& C(w_1,w_2)\left(1 + \hat{Q}(w_1,w_2) +{1\over 2} \left[\hat{Q}(w_1,w_2)\right]^2 + \ldots\right)\rpmket\label{ChiExpansion}
\eea
where the operator $\hat Q$ contains pairs of excitations.  Thus each term in in Equation \ref{ChiExpansion} contains an even number of excitations on the relevant R vacuum.  Inserting this expansion into Equation (\ref{GeneralForm2}), we find:
\bea
\hat \s \, \a^{(i)}_{A\dot A,-m}|\emptyset\rangle &=& \sum_{(j)}\sum_{n} f^{B,(i)(j)}_{mn}\a^{(j)}_{A\dot A,-n}C(w_1,w_2)\left(1 + \hat{Q}(w_1,w_2) + \ldots\right)\rpmket\nn
\hat \s \, d^{(i),\a A}_{-m}|\emptyset\rangle &=& \sum_{(j)}\sum_{n} f^{F\a,(i)(j)}_{mn}d^{(j),\a A}_{-n}C(w_1,w_2)\left(1 + \hat{Q}(w_1,w_2) + \ldots\right)\rpmket,\nn\label{ExpandedGeneralForm}
\eea
where each term now contains an odd number of excitations.

In order to avoid the messy details of the $\hat{Q}$ operator, we should choose a capping state that has a nonzero overlap with only the $0^{\text{th}}$  order term in Equation (\ref{ExpandedGeneralForm}).  We first note that this state should be built upon the same vacuum as $|\chi\rangle$, namely:
\bea
\langle\emptyset'| &\equiv& \rpmbra.
\eea
We will also need a single mode with appropriate charges to contract with the existing excitation from the $0^{\text{th}}$ order term in Equation (\ref{ExpandedGeneralForm}).  This yields, for example:
\bea
\langle\emptyset'|\a^{(j)}_{++,n}\sh\a^{(i)}_{--,-m}|\emptyset\rangle &=& -pf^{B,(i)(j)}_{mn}C(w_1,w_2)\langle\emptyset'|\emptyset'\rangle ~=~ -pf^{B,(i)(j)}_{mn}\langle\emptyset'|\sh|\emptyset\rangle\nn
\langle\emptyset'|d^{(j),++}_n\sh d^{(i),--}_{-m}|\emptyset\rangle &=& -f^{F-,(i)(j)}_{mn}C(w_1,w_2)\langle\emptyset'|\emptyset'\rangle ~=~ -pf^{B,(i)(j)}_{mn}\langle\emptyset'|\sh|\emptyset\rangle\nn
\langle\emptyset'|d^{(j),--}_n\sh d^{(i),++}_{-m}|\emptyset\rangle &=& -f^{F+,(i)(j)}_{mn}C(w_1,w_2)\langle\emptyset'|\emptyset'\rangle ~=~ -pf^{B,(i)(j)}_{mn}\langle\emptyset'|\sh|\emptyset\rangle.\quad\label{CylinderRelations}
\eea
Equivalent relations can be obtained for other choices of the SU(2) indices.  Since we do not here determine the coefficient $C(w_1,w_2)$, we will not make use of the middle expressions.

\subsection{Spectral Flows}
Now that we know the capping states, we can determine the type and location of the spin fields they bring when mapped to the $t$ plane.  Recalling Equation (\ref{ImagePoints}), we have:
\bea
|0_{R,-}\rangle^{(1}) &\to& S^-(t=-a)\nn
|0_{R,-}\rangle^{(2}) &\to& S^-(t=-b)\nn
{}^{(1')}\langle0_{R,+}| &\to& S^-(t=\infty)\nn
{}^{(2')}\langle0_{R,-}| &\to& S^+(t=0).
\eea
We also list the spin fields accompanying the twists.  These are positive spin fields occurring at the bifurcation points in Equation (\ref{BifurcationPoints}):
\bea
\s_2^+(w_1) &\to& S^+\left(t=-\sqrt{ab}\right)\nn
\s_2^+(w_2) &\to& S^+\left(t=\sqrt{ab}\right).
\eea

We must now remove these six spin fields via local spectral flows.  A spectral flow by $\a=+1$ units removes an $S^-$ spin field while also applying an effect at infinity: Either removing an $S^+$ at infinity or applying an $S^-$ at infinity if there is no $S^+$ to be removed.  Table \ref{SpectralFlowTable} tracks the behavior of the $t$ plane both locally and at infinity through five spectral flows.  The net effect is the elimination of all six spin fields in the $t$ plane.  We can now close all punctures with the local NS vacuum.  This vacuum is all that remains at the twist insertions, while the in and out states may carry additional contours from initial and capping excitations.  Thus the coordinate maps in conjunction with our spectral flows produces the following transformations:
\bea
|\emptyset\rangle &\to& \nstket \nn
\langle\emptyset'| &\to& \nstbra \nn
\sh &\to& \one.
\eea
There are also transformations on the bosonic and fermionic modes.  These transformations are the primary focus of the next two sections.  For now, we simply denote the transformed modes with primes:
\bea
\a^{(k)}_{A\dot A,n} &\to& \a'^{(k)}_{A\dot A,n}\nn
d^{(k),\a A}_{n} &\to& d'^{(k),\a A}_{n}.
\eea

\begin{table}[hbt]
\begin{center}
\begin{tabular}{|>{$}l<{$~}l>{~$}l<{$}>{$}c<{$}>{$}l<{$}c>{$}l<{$}|}\hline
\a = -1 & at & t=0 & \to & S^+(t=0) \text{ removed} & \& & S^-(t=\infty) \text{ removed}\\
&&&&&&\\
\a = -1 & at & t=\sqrt{ab} & \to & S^+\left(t=\sqrt{ab}\right) \text{ removed} & \& & S^+(t=\infty) \text{ added}\\
&&&&&&\\
\a = +1 & at & t=-a & \to & S^-(t=-a) \text{ removed} & \& & S^+(t=\infty) \text{ removed}\\
&&&&&&\\
\a = +1 & at & t=-b & \to & S^-(t=-b) \text{ removed} & \& & S^-(t=\infty) \text{ inserted}\\
&&&&&&\\
\a = -1 & at & t=-\sqrt{ab} & \to & S^+\left(t=-\sqrt{ab}\right) \text{ removed} & \& & S^-(t=\infty) \text{ removed}\\
\hline
\end{tabular}
\end{center}
\label{SpectralFlowTable}
\caption{One possible ordering of spectral flows.  While all orders have the same overall effect, we have presented here an order that never brings any complicated operators to the point at infinity.  The net result is the removal of all six spin insertions in the $t$ plane.}
\end{table}

We now apply these transformations to the relations found in Equation (\ref{CylinderRelations}).  Naively, we obtain:
\bea
f^{B,(i)(j)}_{mn} &=& -{1\over n}{\nstbra\a'^{(j)}_{++,n}\a^{(i)}_{--,-m}\nstket\over\nstbra 0_{NS}\rangle_t}\nn
f^{F-,(i)(j)}_{mn} &=& -{\nstbra d'^{(j),++}_nd'^{(i),--}_{-m}\nstket\over\nstbra 0_{NS}\rangle_t} \nn
f^{F+,(i)(j)}_{mn}&=&-{\nstbra d'^{(j),--}_nd'^{(i),++}_{-m}\nstket\over\nstbra 0_{NS}\rangle_t\rangle}.
\eea
At this point there is one more subtlety to consider.  The out state modes on copy 2 are mapped to the origin of the $t$ plane before our in state modes are deformed to this region, so the in modes will have contours wrapped \emph{outside} of the out modes.  This means that we should write all copy 2 out modes to the left of other modes.  For fermions, this order-swapping will also bring a sign change.  Taking all of this into account, we find:
\bea
f^{B,(i)(1')}_{mn} &=& -{1\over n}{\nstbra\a'^{(1')}_{++,n}\a^{(i)}_{--,-m}\nstket\over\nstbra 0_{NS}\rangle_t}\nn
f^{B,(i)(2')}_{mn} &=& -{1\over n}{\nstbra\a^{(i)}_{--,-m}\a'^{(2')}_{++,n}\nstket\over\nstbra 0_{NS}\rangle_t}\nn
f^{F-,(i)(1')}_{mn} &=& -{\nstbra d'^{(1'),++}_nd'^{(i),--}_{-m}\nstket\over\nstbra 0_{NS}\rangle_t} \nn
f^{F-,(i)(2')}_{mn} &=& {\nstbra d'^{(i),--}_{-m}d'^{(2'),++}_n\nstket\over\nstbra 0_{NS}\rangle_t} \nn
f^{F+,(i)(1')}_{mn} &=& -{\nstbra d'^{(1'),--}_nd'^{(i),++}_{-m}\nstket\over\nstbra 0_{NS}\rangle_t} \nn
f^{F+,(i)(2')}_{mn}&=&{\nstbra d'^{(i),++}_{-m}d'^{(2'),--}_n\nstket\over\nstbra 0_{NS}\rangle_t\rangle}.\label{tPlaneRelations}
\eea

In the following two sections we will determine the integral expressions of the transformed modes.  We will then expand these modes in terms of modes natural to the $t$ plane in the appropriate regions.  This allows us to use the commutation relations found in Equations (\ref{bosoncommutation}) and (\ref{fermioncommutation}) to obtain an analytic expression for the transition amplitudes from the above relations.

\section{Calculating $f^B$}\label{Bosons}
Let us recall the expression for the bosonic transition amplitude from Equation (\ref{tPlaneRelations}).
\bea
f^{B,(i)(1')}_{mn} &=& -{1\over n}{\nstbra\a'^{(1')}_{++,n}\a^{(i)}_{--,-m}\nstket\over\nstbra 0_{NS}\rangle_t}\nn
f^{B,(i)(2')}_{mn} &=& -{1\over n}{\nstbra\a^{(i)}_{--,-m}\a'^{(2')}_{++,n}\nstket\over\nstbra 0_{NS}\rangle_t}.\label{BosonfRelation}
\eea

The rest of this session will consist primarily of writing these transformed modes in full detail.  Since bosonic modes are unaffected by spectral flows, we deal only with the coordinate maps.  We first expand the cylinder modes in terms of modes natural to the $t$ plane in the region of their image points.  After this we are free to drop any modes that annihilate the local NS vacuum (non-negative modes in this case).  We then deform the contours from our initial-state modes into the neighborhood of the capping mode.  From here we re-expand the initial-state mode in terms of $t$ plane modes natural to the new neighborhood.  We can then apply the commutation relations from Equation (\ref{bosoncommutation}) to obtain an expression for the transition amplitude. 

\subsection{The Boson Mode Expansions}
In \cite{chm1}, the transformed final modes were presented in full detail.  Here we merely give those results:
\bea
\a'^{(1')}_{A\dot A,n} &=& \sum_{j,j' = 0}^{\infty} {}^n C_j {}^n C_{j'} a^j b^{j'} \tilde{\a}_{A\dot A,n-j-j'}^{t\to\infty} \\
\a'^{(2')}_{A\dot A,n} &=& -\sum_{j,j' = 0}^{\infty} {}^n C_j {}^n C_{j'} a^{n-j} b^{n-j'} \tilde{\a}_{A\dot A,j+j'-n}^{t\to0}.
\eea
where ${}^n C_k={n!\over k!(n-k)!}$ is the Binomial coefficient. 

In keeping with the notation of \cite{chm1} we chose to have the initial Copy 1 map to the neighborhood around $t=a$ while the initial Copy 2 maps to the neighborhood around $t=b$.  Since the combination $\partial X(w)\diff w$ has weight zero under coordinate transformations, we need only rewrite the $e^{nw}$ factor in terms of $t$.  Following Equations (\ref{zMap}) and (\ref{tMap}), this is:
\bea
e^{nw} &=& z^n ~=~ \left( {(t+a)(t+b)\over t}\right)^n.
\eea
The initial boson modes are then:
\bea
\a'^{(1)}_{A\dot A,-m} &=& {1\over 2\pi} \oint_{t=-a} X_{A\dot A}(t) \left ( {(t+a)(t+b) \over t} \right )^{-m} \diff t\nn
&=& {1\over 2\pi} \oint_{t'=0} X_{A\dot A}(t) \left ( {t'(t'-a+b) \over t'-a} \right )^m \diff t' \label{1}\\
\a'^{(2)}_{A\dot A,-m} &=& {1\over 2\pi} \oint_{t=-b} X_{A\dot A}(t') \left ( {(t+a)(t+b) \over t} \right )^{-m} \diff t\nn
&=&{1\over 2\pi} \oint_{t''=0} X_{A\dot A}(t'') \left ( {(t''+a-b)t'' \over t''-b} \right )^{-m} \diff t'', \label{2}
\eea
where we have introduced shifted coordinates $t' = t-a$ and $ t'' = t-b$.  We shall deal with each copy in turn.

\subsubsection*{Copy 1 Initial Modes}
We now expand the integrand of (\ref{1}) in powers of $t'$.
\bea
\left ( {t'(t'-a+b) \over t'-a} \right )^{-m} &=& t'^{-m} \left ( t' + (b-a)\right )^{-m} (t'-a)^m\nn
&=& t'^{-m} \sum_{k=0}^{\infty} {}^{-m} C_k t'^k (b-a)^{-m-k} \sum_{k'=0}^{\infty} {}^{m} C_{k'} t'^{k'} (-a)^{m-k'}\nn
&=& \sum_{k,k' = 0}^{\infty} {}^{-m} C_k {}^{m}C_{k'} (b-a)^{-m-k} (-a)^{m-k'}t'^{k+k'-m} \nn
&\to&\sum_{k,k' = 0}^{k+k'<m} {}^{-m} C_k {}^{m}C_{k'} (b-a)^{-m-k} (-a)^{m-k'}t'^{k+k'-m}.  \label{1middle}
\eea
Here we have placed a limit on the sum to ensure that we work only with creation operators.  This means we have dropped those modes which annihilate the local NS vacuum.

We must now bring the contour to one of two different regions, depending on which copy of final modes we wish to interact with.  When interacting with Copy 1 final modes we must bring the contour out to infinity, thus expanding (\ref{1middle}) around $t=\infty$.  This will leave our contour \emph{inside} the final mode contour, so we expand in terms of creation operators.  When interacting with Copy 2 final modes we must instead bring the contour to the origin of the $t$ plane, thus expanding (\ref{1middle}) around $t=0$.  This time our contour is \emph{outside} the final mode contour (indeed smaller $|z|$ map to larger $|t|$ in this region).  We must thus expand this region in terms of annihilation operators.  We also pick up an extra minus sign in this region to account from the contour reversing direction when it wraps around a finite point (the $t$ plane origin).

When expanding the Copy 1 initial modes for interaction with Copy 1 final modes, we find:
\bea
\left[ t'^{k+k'-m}\right]_{t \to \infty} &=& (t+a)^{k+k'-m}\nn
&=& t^{k+k'-m}\left(1+{a\over t}\right)^{k+k'-m}\nn
&=& t^{k+k'-m}\sum_{k''=0}^{\infty} {}^{k+k'-m} C_{k''}a^{k''}t^{-k''}\nn
&=& \sum_{k''=0}^{\infty} {}^{k+k'-m} C_{k''}a^{k''}t^{k+k'-m-k''},
\eea
where ${}^n C_m$ is the binomial coefficient of $m$ and $n$.
Plugging this into Equation (\ref{1middle}) gives:
\bea
\left ( {t'(t'-a+b) \over t'-a} \right )^{-m}_{t\to \infty}&\!\!\!=&\!\!\!\!\sum_{k,k',k'' = 0}^{\infty}\!\!\! {}^{-m}C_k {}^{m}C_{k'} {}^{k+k'-m} C_{k''} t^{k+k'-m-k''} \nn
&&\qquad {}\times(-1)^{m-k'}(b-a)^{-m-k} (a)^{m+k''-k'},
\eea
and thus:
\bea
\a'^{(1)}_{A\dot A,-m} \!\!&=& \!\!\!\!\sum_{k,k',k'' = 0}^{\infty} \!\!{}^{-m}C_k {}^{m}C_{k'} {}^{k+k'-m} C_{k''}(-1)^{m-k'}(b-a)^{-m-k} a^{m+k''-k'} \tilde{\a}_{A\dot A,k+k'-m-k''}^{t\to\infty},\nn \label{B1InitialLarge}
\eea
where we still have the constraint $k+k'<m$.

For interaction with the Copy 2 final modes, we instead have:
\bea
\left[t'^{k+k'-m}\right]_{t \to 0} &=& (t+a)^{k+k'-m}\nn
&=& a^{k+k'-m}\left(1+{t\over a}\right)^{k+k'-m}\nn
&=& a^{k+k'-m}\sum_{k''=0}^{\infty} {}^{k+k'-m} C_{k''}a^{-k''}t^{k''}\nn
&=& \sum_{k''=0}^{\infty} {}^{k+k'-m} C_{k''}a^{k+k'-m-k''}t^{k''}.
\eea
This gives:
\bea
\left ( {t'(t'-a+b) \over t'-a} \right )^{-m}_{t\to \infty}&\!\!\!\!=&\!\!\!\!\sum_{k,k',k'' = 0}^{\infty}\!\! {}^{-m}C_k {}^{m}C_{k'} {}^{k+k'-m} C_{k''}(-1)^{m-k'}(b-a)^{-m-k} (a)^{k-k''}t^{k''},\nn
\eea
and thus:
\bea
\a'^{(1)}_{A\dot A,-m} \!\!&=& \!\!\!\!-\sum_{k,k',k'' = 0}^{\infty}\!\!{}^{-m}C_k {}^{m}C_{k'}{}^{k+k'-m} C_{k''}(-1)^{m-k'}(b-a)^{-m-k} a^{k-k''}\tilde{\a}_{A\dot A,k''}^{t\to0}, \label{B1InitialSmall}
\eea
where again $k+k'<m$ and we have an extra minus sign from the change in direction of the contour.

\subsubsection*{Copy 2 Initial Modes}
In principle, we now expand the integrand of (\ref{2}) in powers of $t''$.  In practice we can shortcut this by noticing that this integrand can be obtained from integrand of (\ref{1}) via the interchange $a \leftrightarrow b$, under which the reparameterization variables are also interchanged, $t' \leftrightarrow t''$.  In other words, $a \leftrightarrow b$ amounts to a redefinition of our initial copies $(1)\leftrightarrow(2)$.  We can thus apply $a \leftrightarrow b$ to (\ref{B1InitialLarge}) and (\ref{B1InitialSmall}), yielding:
\bea
\a'^{(2)}_{A\dot A,-m} \!\!&=&\!\!\!\! \sum_{k,k',k''=0}^{\infty}\!\! {}^{-m}C_k {}^m C_{k'} {}^{k+k'-m}C_{k''}(-1)^{m-k}(a-b)^{-m-k}b^{m+k''-k'}\tilde{\a}_{A\dot A,k+k'-m-k''}^{t\to\infty}\label{B2InitialLarge}\nn
\a'^{(2)}_{A\dot A,-m} \!\!&=&\!\!\!\! -\sum_{k,k',k''=0}^{\infty}\!\!{}^{-m}C_k {}^m C_{k'}{}^{k+k'-m}C_{k''}(-1)^{m-k}(a-b)^{-m-k}b^{k-k''}\tilde{\a}_{A\dot A,k''}^{t\to0}\label{B2InitialSmall},
\eea
with $k+k'<m$.

\subsubsection*{Summary}
For ease of reference, we group all of the expansions together.
\bea
\a'^{(1)}_{A\dot A,-m} \!\!&=& \!\!\!\!\sum_{k,k',k'' = 0}^{\infty} \!\!{}^{-m}C_k {}^{m}C_{k'} {}^{k+k'-m} C_{k''}(-1)^{m-k'}(b-a)^{-m-k} a^{m+k''-k'} \tilde{\a}_{A\dot A,k+k'-m-k''}^{t\to\infty}\nn
\!\!&=& \!\!\!\!-\sum_{k,k',k'' = 0}^{\infty} \!\!{}^{-m}C_k {}^{m}C_{k'} {}^{k+k'-m} C_{k''}(-1)^{m-k'}(b-a)^{-m-k} a^{k-k''} \tilde{\a}_{A\dot A,k''}^{t\to0}\label{1i}
\eea
\bea
\a'^{(2)}_{A\dot A,-m} \!\!&=&\!\!\!\! \sum_{k,k',k''=0}^{\infty}\!\! {}^{-m}C_k {}^m C_{k'} {}^{k+k'-m}C_{k''}(-1)^{m-k}(a-b)^{-m-k}b^{m+k''-k'}\tilde{\a}_{A\dot A,k+k'-m-k''}^{t\to\infty}\nn
\!\!&=&\!\!\!\!-\sum_{k,k',k''=0}^{\infty}\!\!{}^{-m}C_k {}^m C_{k'} {}^{k+k'-m}C_{k''}(-1)^{m-k}(a-b)^{-m-k}b^{k-k''}\tilde{\a}_{A\dot A,k''}^{t\to0}\label{2i}
\eea
\bea
\a'^{(1')}_{A\dot A,n} &=& \sum_{j,j' = 0}^{\infty} {}^n C_j {}^n C_{j'} a^j b^{j'} \tilde{\a}_{A\dot A,n-j-j'}^{t\to\infty} \label{1f}
\eea
\bea
\a'^{(2')}_{A\dot A,n} &=& -\sum_{j,j' = 0}^{\infty} {}^n C_j {}^n C_{j'} a^{n-j} b^{n-j'} \tilde{\a}_{A\dot A,j+j'-n}^{t\to0}\label{2f}.
\eea
For all initial-mode cases we have the additional constraint:
\bea
k+k' < m.
\eea

\subsection{Computing the $f^B$ Coefficients}
We can now compute the $f^B$ coefficients by plugging (\ref{1i} $-$ \ref{2f}) into (\ref{BosonfRelation}).  We'll split the cases into subsections.

\subsubsection{$(i)=(j)=1$}
Using the first line of (\ref{1i}) along with (\ref{1f}) and (\ref{BosonfRelation}), we find:
\bea
f^{B,(1)(1')}_{m,n} &=& \sum_{j,j',k,k',k''=0}^{\infty}{}^n C_j {}^n C_{j'}{}^{-m}C_k{}^m C_{k'}{}^{k+k'-m}C_{k''}(-1)^{m-k'}a^{m+k''-k'+j}b^{j'}(b-a)^{-m-k} \nn
&&{}\times{}\left (- {1\over n} {{}_t \langle 0|\tilde{\a}_{++,n-j-j'}^{t\to\infty}\tilde{\a}_{--,k+k'-m-k''}^{t\to\infty}|0\rangle_t\over {}_t \langle 0|0\rangle_t} \right ),
\eea
with the constraint:
\bea
k+k'<m.
\eea
We now apply the appropriate commutation relation from Equation (\ref{bosoncommutation}).  The amplitude ratio gives zero unless the following constraints are met:
\bea
n-j-j' > 0 &\implies& j+j' < n \label{c1}\\
k+k'-m-k'' < 0 &\implies& k+k' < m+k'' \geq m \label{c2}\\
n-j-j' = -(k+k'-m-k'') &\implies& k'' = n+k+k'-m-j-j'\geq 0.\label{c3}
\eea
When nonzero the commutator is simply $-(n-j-j')$.

Let us now look at our constraints in more detail.  Equation (\ref{c2}) is redundant with the earlier constraint $k+k' < m$.  Meanwhile, the right side of Equation (\ref{c3}) gives:
\bea
k'' = n+k+k'-m-j-j'\geq 0 &\implies& j+j' \leq n - m + k + k',
\eea
which is stricter than (\ref{c1}) since $m>0$.  From here, the fact that $j+j'\geq 0$ gives:
\bea
0\leq j+j' \leq n - m + k + k') &\implies& m-n \leq k+k',
\eea
which places another \emph{lower} bound on $k'$.  We can now eliminate the $k''$ sum and use the constraints to set the limits.
\bea
\max(m-n-k,0) &\leq ~k'~ \leq& m-k-1 \nn
0 &\leq ~k~ \leq& m-1 \nn
0 &\leq ~j'~ \leq& n-(m-k-k')-j \nn
0 &\leq ~j~ \leq& n-(m-k-k'). \label{j=1limits}
\eea
This gives:
\bea
f^{B,(1)(1')}_{m,n} &\!\!=&\!\! \sum_{k=0}^{m-1}\,\sum_{k'=\max(m-n-k,0)}^{m-k-1}\,\sum_{j=0}^{n-m+k+k'}\,\sum_{j'=0}^{n-m+k+k'-j}{}^n C_j {}^n C_{j'} {}^{-m}C_k {}^m C_{k'}\nn
&&\quad{}\times {}^{k+k'-m}C_{n+k+k'-m-j-j'}{n-j-j' \over n}(-1)^{m-k'}a^{n+k-j'}b^{j'}(b-a)^{-m-k}.\qquad\label{f11}
\eea

\subsubsection{$(i)=1,(j)=2$}
Here we use the second line of (\ref{1i}) along with (\ref{1f}) and (\ref{BosonfRelation}).  Noting that the initial mode maps to a contour \emph{outside} of the final mode due to $z \sim t^{-1}$, we find:
\bea
f^{B,(1)(2')}_{m,n} &=& \sum_{j,j',k,k',k''} {}^n C_j {}^n C_{j'}{}^{-m}C_k {}^m C_{k'} {}^{k+k'-m}C_{k''} (-1)^{m-k'} a^{k-k''+n-j} b^{n-j'}(b-a)^{-m-k} \nn
&&\quad{}\times{}\left (-{1\over n}{{}_t\langle 0|\tilde{a}_{--,k''}^{t\to0}\tilde{a}_{++, j+j'-n}^{t\to0}|0\rangle_t \over {}_t \langle 0| 0 \rangle_t} \right ),
\eea
with $k+k' < m$.  Here we again have constraints for the amplitude ratio to remain nonzero:
\bea
j+j'-n < 0 &\implies& j+j' < n \\
k'' = n-j-j' &\implies& \text{No extra limits}
\eea
When these conditions are met, the amplitude ratio gives a factor:
\bea
-k'' = -(n-j-j')
\eea
We can now eliminate the $k''$ sum and use the constraints to set the limits:
\bea
&0 \leq k' \leq m-k-1 \nn
&0 \leq k \leq m-1 \nn
&0 \leq j' \leq n-j-1 \nn
&0\leq j \leq n-1 \label{j=2limits}
\eea
\bea
f^{B,(1)(2')}_{m,n} &\!\!=& \!\!\sum_{k=0}^{m-1}\,\sum_{k'=0}^{m-k-1}\,\sum_{j=0}^{n-1}\,\sum_{j'=0}^{n-j-1}{}^n C_j {}^n C_{j'}{}^{-m}C_k {}^m C_{k'} {}^{k+k'-m}C_{n-j-j'}\nn
&&\qquad\qquad\qquad\qquad{}\times{}{n-j-j' \over n}(-1)^{m-k'}a^{k+j'} b^{n-j'}(b-a)^{-m-k}\label{f12}
\eea

\subsubsection{The Other Cases by Symmetry}
The other cases can be obtained easily from these first two by applying the interchange $a \leftrightarrow b$, which swaps the initial copies while leaving the final copies unchanged.  We thus find:
\bea
f^{B,(2)(1')}_{m,n} &=& \left [ f^{B,(1)(1')}_{m,n}\right ]_{a\leftrightarrow b} \nn
&\!\!=&\!\! \sum_{k=0}^{m-1}\,\sum_{k'=\max(m-n-k,0)}^{m-k-1}\,\sum_{j=0}^{n-m+k+k'}\,\sum_{j'=0}^{n-m+k+k'-j}{}^n C_j {}^n C_{j'} {}^{-m}C_k {}^m C_{k'}\nn
&&\quad{}\times {}^{k+k'-m}C_{n+k+k'-m-j-j'}{n-j-j' \over n}(-1)^{m-k'}b^{n+k-j'}a^{j'}(a-b)^{-m-k}\qquad\label{f21}\\\nn
f^{B,(2)(2')}_{m,n} &=& \left [ f^{B,(1)(2')}_{m,n} \right ]_{a\leftrightarrow b} \nn
&=& \!\!\sum_{k=0}^{m-1}\,\sum_{k'=0}^{m-k-1}\,\sum_{j=0}^{n-1}\,\sum_{j'=0}^{n-j-1}{}^n C_j {}^n C_{j'}{}^{-m}C_k {}^m C_{k'} {}^{k+k'-m}C_{n-j-j'}\nn
&&\qquad\qquad\qquad\qquad{}\times{}{n-j-j' \over n}(-1)^{m-k'}a^{n-j'}b^{k+j'}(a-b)^{-m-k}.\label{f22}
\eea

\subsection{Results}
Here we gather the results together for convenient reference.
\bea
f^{B,(1)(1')}_{m,n} &\!\!=&\!\! \sum_{k=0}^{m-1}\,\sum_{k'=\max(m-n-k,0)}^{m-k-1}\,\sum_{j=0}^{n-m+k+k'}\,\sum_{j'=0}^{n-m+k+k'-j}{}^n C_j {}^n C_{j'} {}^{-m}C_k {}^m C_{k'}\nn
&&\quad{}\times {}^{k+k'-m}C_{n+k+k'-m-j-j'}{n-j-j' \over n}(-1)^{m-k'}a^{n+k-j'}b^{j'}(b-a)^{-m-k}\qquad\\\nn
f^{B,(1)(2')}_{m,n} &\!\!=& \!\!\sum_{k=0}^{m-1}\,\sum_{k'=0}^{m-k-1}\,\sum_{j=0}^{n-1}\,\sum_{j'=0}^{n-j-1} (-1)^{m-k'}{n-j-j' \over n} a^{k+j'} b^{n-j'}(b-a)^{-m-k}\nn
&&\qquad\qquad\qquad\qquad{}\times{}^n C_j {}^n C_{j'}{}^{-m}C_k {}^m C_{k'} {}^{k+k'-m}C_{n-j-j'}\\\cr
f^{B,(2)(1')}_{m,n} &\!\!=&\!\! \sum_{k=0}^{m-1}\,\sum_{k'=\max(m-n-k,0)}^{m-k-1}\,\sum_{j=0}^{n-m+k+k'}\,\sum_{j'=0}^{n-m+k+k'-j}{}^n C_j {}^n C_{j'} {}^{-m}C_k {}^m C_{k'}\nn
&&\quad{}\times {}^{k+k'-m}C_{n+k+k'-m-j-j'}{n-j-j' \over n}(-1)^{m-k'}b^{n+k-j'}a^{j'}(a-b)^{-m-k}\qquad\\\cr
f^{B,(2)(2')}_{m,n} &\!\!=&\!\!\sum_{k=0}^{m-1}\,\sum_{k'=0}^{m-k-1}\,\sum_{j=0}^{n-1}\,\sum_{j'=0}^{n-j-1} (-1)^{m-k'}{n-j-j' \over n} a^{n-j'}b^{k+j'}(a-b)^{-m-k}\nn
&&\qquad\qquad\qquad\qquad{}\times{}^n C_j {}^n C_{j'}{}^{-m}C_k {}^m C_{k'} {}^{k+k'-m}C_{n-j-j'}.
\eea
Using the relationships from Equations (\ref{NegativeRelation}) and (\ref{PositiveRelation}), this also gives the fermion transition amplitude for all cases that do not involve a zero mode.

\section{Fermion Zero Modes}\label{Fermions}

We have seen that the $f$ functions for fermions can be related to the $f$ functions for bosons. These relation however do not hold when one or more of the modes involved in the relation is a fermion zero mode. In this case we need to be more careful, and compute the $f$ functions for fermions explicitly. In this section we will perform the relevant computations for the $f^{F(i)(j),\pm}$ when one or more of the indices represent a zero mode. .

We follow the same method used for the bosons, except that we must now account for the effects of our five spectral flows. These spectral flows were required to map the originital problem to a plane with no punctures. The bosons were not affected by the spectral flow, but the fermions will be. The full calculation of the spectral flow effects was performed in  \cite{chm1} for capping modes, so we present only the results found there:
\bea
d'^{(1'),+A}_{n}&=&\sum_{j,j'\geq 0} {}^{n-1}C_j{}^{n-1}C_{j'} a^j b^{j'} \tilde{d}^{+A,t\to\infty}_{n-j-j'+1/2}\nn
&&{}-\sum_{j,j'\geq 0} {}^{n-1}C_j{}^{n-1}C_{j'} a^{j+1} b^{j'+1}\tilde{d}_{n-j-j'-3/2}^{+A,t\to\infty}\nn
d'^{(1'),-A}_{n}&=&\sum_{j,j'\geq 0} {}^{n}C_j{}^{n}C_{j'} a^{j} b^{j'}\tilde{d}^{-A,t\to\infty}_{n-j-j'-1/2}\nn
d'^{(2'),+A}_{n}&=&-\sum_{j,j'\geq 0}{}^{n-1}C_j{}^{n-1}C_{j'} a^{n-j-1} b^{n-j'-1} \tilde{d}^{+A,t\to0}_{j+j'-n+5/2}\nn
&&{}+\sum_{j,j'\geq 0} {}^{n-1}C_j{}^{n-1}C_{j'} a^{n-j} b^{n-j'}\tilde{d}^{+A,t\to0}_{j+j'-n+1/2}\nn
d'^{(2'),-A}_{n}&=&-\sum_{j,j'\geq 0} {}^{n}C_j{}^{n}C_{j'} a^{n-j} b^{n-j'}\tilde{d}^{-A,t\to0}_{j+j'-n-1/2}
\label{tplanemodesNS}
\eea
Since we are interested in the cases containing at least one zero mode, we evaluate (\ref{tplanemodesNS}) for such cases.
\bea
d'^{(1'),+A}_{0}&=&\sum_{j,j'\geq 0}(-1)^{j+j'} a^j b^{j'} \tilde{d}^{+A,t\to\infty}_{-j-j'+1/2}-\sum_{j,j'\geq 0} (-1)^{j+j'} a^{j+1} b^{j'+1}\tilde{d}_{-j-j'-3/2}^{+A,t\to\infty}\nn
d'^{(1'),-A}_{0}&=&\tilde{d}^{-A,t\to\infty}_{-1/2}\nn
d'^{(2'),+A}_{0}&=&-\sum_{j,j'\geq 0}(-1)^{j+j'} a^{-j-1} b^{-j'-1} \tilde{d}^{+A,t\to0}_{j+j'+5/2}+\sum_{j,j'\geq 0}(-1)^{j+j'} a^{-j} b^{-j'}\tilde{d}^{+A,t\to0}_{j+j'+1/2}\nn
d'^{(2'),-A}_{0}&=&-\tilde{d}^{-A,t\to0}_{-1/2}
\label{tplanezeromodesNS}
\eea

We now turn to modes of the in state.  We will work in three subsections.  First, we will follow the in states through the coordinate shifts and spectral flows.  Next we will express the transformed modes in terms of modes natural to the $t$ plane in the appropriate neighborhood, stretch the contours to the neighborhood of the out states, and then re-express the result in modes natural to the new neighborhoods.  Lastly we will plug the results into Equation (\ref{tPlaneRelations}) to obtain expressions for the fermion transition amplitudes.

\subsection{Coordinate maps and spectral flows}
We begin with the fermion modes on the cylinder before the twist insertions:
\bea
d^{(i),\a A}_n &=& {1\over 2\pi i}\int_{\t<\t_1,\s=0}^{\s=2\pi} \psi^{(i),\a A}(w)e^{nw}\diff w\nn
d^{(j),\a A}_n &=& {1\over 2\pi i}\int_{\t>\t_2,\s=0}^{\s=2\pi} \psi^{(j),\a A}(w)e^{nw}\diff w.\label{InModesCylinder}
\eea
The fermion field $\p$ has conformal weight $1/2$.  Combined with the jacobian, this gives:
\bea
\p^{(i),\a A}(x)\diff x &\to& \left({\diff x' \over \diff x}\right)^{-\h}\p^{(i),\a A}(x')\diff x'.
\eea

The derivatives of our two maps are as follows:
\bea
{\diff z \over \diff w} &=& e^w ~=~ z\\\nn
{\diff t \over \diff z} &=& \left({\diff z \over \diff t}\right)^{-1}\nn
&=& \left(1-{ab\over t^2}\right)^{-1}\nn
&=& t^2\left(t^2-ab\right)^{-1}.
\eea
Taking the two coordinate shifts together, we find:
\bea
\psi^{(i),\a A}(w)e^{nw}\diff w &\to& z^{n-\h}t^{-1}\left(t^2-ab\right)^{\h}\psi^{(i),\a A}(t)\diff t\label{FermionCoordinateResult}\\
&=& (t+a)^{n-\h}(t+b)^{n-\h}t^{-n-\h}\left(t^2-ab\right)^{\h}\psi^{(i),\a A}(t)\diff t.\nonumber
\eea

We must now apply our five spectral flows.  For each spectral flow by a unit $\a$ around a point $t_0$, a field with an SU(2) R charge q transforms as:
\bea
\psi^q(t) &\to& (t-t_0)^{-\a q}\psi^q(t).
\eea
We can determine the factors that come from the spectral flows outlined in Table (\ref{SpectralFlowTable}).  These factors are presented in Table (\ref{SFResultsTable}).  Applying this to Equation (\ref{FermionCoordinateResult}) gives:
\bea
\psi^{(i),+A}(w)e^{nw}\diff w &\to& (t+a)^{n}(t+b)^{n-1}t^{-n-1}\left(t^2 - ab\right)\psi^{(i),+A}(t)\diff t\nn
\psi^{(i),-A}(w)e^{nw}\diff w &\to& t^{-n-1}(t+a)^{n}(t+b)^{n}\psi^{(i),+A}(t)\diff t.\label{FermionTransformations}
\eea

\begin{table}[hbt]
\begin{center}
\begin{tabular}{|>{$}c<{$~}|>{$}c<{$~}|>{~$}l<{$\quad}|>{$}l<{$}|}\hline
\a & t_0 & \psi^{+A}\text{ factor} & \psi^{-A}\text{ factor}\\
\hline
-1 & 0 & t^{\h} & t^{-\h}\\
-1 & \sqrt{ab} & \left(t-\sqrt{ab}\right)^{\h} & \left(t-\sqrt{ab}\right)^{-\h}\\
+1 & -a & (t+a)^{-\h} & (t+a)^{\h}\\
+1 & -b & (t+b)^{-\h} & (t+b)^{\h}\\
-1 & -\sqrt{ab} & \left(t+\sqrt{ab}\right)^{\h} & \left(t+\sqrt{ab}\right)^{-\h}\\
\hline
\text{All} && t^{\h}(t+a)^{-\h}(t+b)^{-\h}\left(t^2-ab\right)^{\h} & t^{-\h}(t+a)^{\h}(t+b)^{\h}\left(t^2-ab\right)^{-\h}\\
\hline
\end{tabular}
\end{center}
\label{SFResultsTable}
\caption{The factors obtained by the five spectral flows for the two possible fermion field charges. The last row groups all the factors together in a simplified form.}
\end{table}

Plugging Equation (\ref{FermionTransformations}) into (\ref{InModesCylinder}), we find:
\begin{eqnarray}
d^{(1),+A}_{n}&\to&\frac{1}{2\pi i}\oint_{t'=0}   
\psi^{+A}(t')((t' - a)^{2}-ab)(t'-a)^{-n}t'^{n-1}(t'+b-a)^{n-1}\diff t'\cr
d^{(1),-A}_{n}&\to&\frac{1}{2\pi i}\oint_{t'=0}  
\psi^{-A}(t')(t'-a)^{-n-1}t'^n(t'+b-a)^{n}\diff t'\cr
d^{(2),+A}_{n}&\to&\frac{1}{2\pi i}\oint_{t''=0}  
\psi^{+A}(t'')((t'' - b)^{2}-ab)(t''-b)^{-n}t''^{n-1}(t''+a-b)^{n-1}\diff t''\cr
d^{(2),-A}_{n}&\to&\frac{1}{2\pi i}\oint_{t''=0}  
\psi^{-A}(t'')(t''-b)^{-n-1}t''^n(t''+a-b)^{n}\diff t''.\label{tr_sf_modes}
\end{eqnarray}
Here we use the same notation as for the bosons, $t' = t-a$ and $t'' = t-b$.

\subsection{Natural modes on the $t$ plane}
We now expand the in state modes in terms of modes natural to the $t$-plane.  For both $+$ and $-$ cases,  we perform expansions first around $t'=0$ or $t''=0$ and then around $t=\infty$ or $t=0$.  In each case, we denote the integrand from Equation (\ref{tr_sf_modes}) as $I$ with appropriate indices.  We will also make use of the $a \leftrightarrow b$ interchange to swap the initial copies as we did for the bosons.

\subsubsection*{\underline{$d_{-m}^{(i),+A}$}}
\bea
I^{(1),+}&=&((t' - a)^{2}-ab)(t'-a)^{-n}t'^{n-1}(t'+b-a)^{n-1}\cr\cr
&=&(t'-a)^{m+2}t'^{-m-1}(t'+(b-a))^{-m-1}-ab(t'-a)^{m}t'^{-m-1}(t'+(b-a))^{-m-1}\cr\cr
&=&(-a)^{m+2}(b-a)^{-m-1}\left(1+\frac{t'}{-a}\right)^{m+2}t'^{-m-1}\left(1+\frac{t'}{b-a}\right)^{-m-1}\nn\nn
&&\qquad-ab(-a)^{m}(b-a)^{-m-1}\left(1+\frac{t'}{-a}\right)^{m}t'^{-m-1}\left(1+\frac{t'}{b-a}\right)^{-m-1}\nn\nn
&=&\sum_{k,k'\geq 0}{}^{m+2}C_{k}{}^{-m-1}C_{k'}(-a)^{m-k+2}(b-a)^{-m-k'-1}t'^{-m+k+k'-1}\nn
&&\qquad- \sum_{k,k'\geq 0}{}^{m}C_{k}{}^{-m-1}C_{k'}(-1)^{m-k}b\,a^{m-k+1}(b-a)^{-m-k'-1}t'^{-m+k+k'-1}\nn
&=& \sum_{k,k'\geq0}{}^{-m-1}C_{k'}\left({}^{m+2}C_{k}a-{}^{m}C_{k}b\right)(-1)^{m-k}a^{m-k+1}(b-a)^{-m-k'-1}t'^{-m+k+k'-1}\nn\nn
I^{(2),+}&=&I^{(1),+}\left(a\leftrightarrow b, t'\leftrightarrow t''\right)
\label{Integrand+}
\eea

\subsubsection*{\underline{$d_{-m}^{(i),-A}$}}
\bea
I^{(1),-}&=&(t'-a)^{m-1}t'^{-m}(t'+(b-a))^{-m}\nn
&=&(-a)^{m-1}(b-a)^{-m}\left(1+\frac{t'}{-a}\right)^{m-1}t'^{-m}\left(1+\frac{t'}{b-a}\right)^{-m}\nn
&=&\sum_{k,k'\geq 0}{}^{m-1}C_{k}{}^{-m}C_{k'}(-a)^{m-k-1}(b-a)^{-m-k'}t'^{-m+k+k'}\nn\nn
I^{(2),-}&=& I^{(1),-}\left(a\leftrightarrow b,t'\leftrightarrow t''\right)
\label{Integrand-}
\eea

We now insert these expansions back into Equation (\ref{tr_sf_modes}) and drop any powers of $t$ that give local annihilators.  This leaves us with:
\bea
d^{(1),+A}_{-m}&\!\!\to\!\!&\sum_{k=0}^{m}\sum_{k'=0}^{m-k}{}^{-m-1}C_{k'}\left({}^{m+2}C_{k}a-{}^{m}C_{k}b\right)(-1)^{m-k}a^{m-k+1}(b-a)^{-m-k'-1}\tilde{d}^{+A,t\to a}_{-m+k+k'-1/2}\nn
d^{(2),+A}_{-m}&\!\!\to\!\!&\sum_{k=0}^{m}\sum_{k'=0}^{m-k}{}^{-m-1}C_{k'}\left({}^{m+2}C_{k}b-{}^{m}C_{k}a\right)(-1)^{m-k}b^{m-k+1}(a-b)^{-m-k'-1}\tilde{d}^{+A,t\to b}_{-m+k+k'-1/2}\nn
d^{(1),-A}_{-m}&\!\!\to\!\!&\sum_{k=0}^{m-1}\sum_{k'=0}^{m-k-1}{}^{m-1}C_{k}{}^{-m}C_{k'}(-1)^{m-k-1}a^{m-k-1}(b-a)^{-m-k'}\tilde{d}^{-A,t\to a}_{-m+k+k'+1/2}\nn
d^{(2)i,-A}_{-m}&\!\!\to\!\!&\sum_{k=0}^{m-1}\sum_{k'=0}^{m-k-1}{}^{m-1}C_{k}{}^{-m}C_{k'}(-1)^{m-k-1}b^{m-k-1}(a-b)^{-m-k'}\tilde{d}^{-A,t\to b}_{-m+k+k'+1/2}
\label{initialmodestplane}
\eea

We must now expand these modes around $t=\infty$ and $t=0$.  
We do this by simply expanding the integrands hidden in the modes of (\ref{initialmodestplane}) in these regions. 
\subsubsection*{\underline{$t=\infty$}}
\bea
\tilde{d}^{(1)+A,t\to a}_{-m+k+k'-\h}&:&t'^{-m+k+k'-1}=(t+a)^{-m+k+k'-1}=\sum_{k''\geq 0}{}^{-m+k+k'-1}C_{k''}a^{k''}t^{-m+k+k'-k''-1}\nn
\tilde{d}^{(2)+A,t\to b}_{-m+k+k'-\h}&:&t''^{-m+k+k'-1}=(t+b)^{-m+k+k'-1}=\sum_{k''\geq 0}{}^{-m+k+k'-1}C_{k''}b^{k''}t^{-m+k+k'-k''-1}\nn
\tilde{d}^{(1)-A,t\to a}_{-m+k+k'+\h}&:&t'^{-m+k+k'}=(t+a)^{-m+k+k'}=\sum_{k''\geq 0}{}^{-m+k+k'}C_{k''}a^{k''}t^{-m+k+k'-k''}\nn
\tilde{d}^{(2)-A,t\to b}_{-m+k+k'+\h}&:&t''^{-m+k+k'}=(t+b)^{-m+k+k'}=\sum_{k''\geq 0}{}^{-m+k+k'}C_{k''}b^{k''}t^{-m+k+k'-k''}.
\eea
\subsubsection*{\underline{$t=0$}}
\bea
\tilde{d}^{(1),+A,t\to a}_{-m+k+k'-\h}&:&t'^{-m+k+k'-1}=(t+a)^{-m+k+k'-1}=\sum_{k''\geq 0}{}^{-m+k+k'-1}C_{k''}a^{-m+k+k'-k''-1}t^{k''}\nn
\tilde{d}^{(2),+A,t\to b}_{-m+k+k'-\h}&:&t''^{-m+k+k'-1}=(t+b)^{-m+k+k'-1}=\sum_{k''\geq 0}{}^{-m+k+k'-1}C_{k''}b^{-m+k+k'-k''-1}t^{k''}\nn
\tilde{d}^{(1),-A,t\to a}_{-m+k+k'+\h}&:&t'^{-m+k+k'}=(t+a)^{-m+k+k'}=\sum_{k''\geq 0}{}^{-m+k+k'}C_{k''}a^{-m+k+k'-k''}t^{k''}\nn
\tilde{d}^{(2),-A,t\to b}_{-m+k+k'+\h}&:&t''^{-m+k+k'}=(t+b)^{-m+k+k'}=\sum_{k''\geq 0}{}^{-m+k+k'}C_{k''}b^{-m+k+k'-k''}t^{k''}.
\eea
Equation (\ref{initialmodestplane}) now gives:
\subsubsection*{\underline{$t=\infty$}}
\bea
d'^{(1),+A}_{-m}&=&\sum_{k=0}^{m}\sum_{k'=0}^{m-k}\sum_{k''\geq0}{}^{-m-1}C_{k'}{}^{-m+k+k'-1}C_{k''}\left({}^{m+2}C_{k}a-{}^{m}C_{k}b\right)\nn
&&\quad{}\times(-1)^{m-k}a^{m-k+k''+1}(b-a)^{-m-k'-1}\tilde{d}^{+A,t\to\infty}_{-m+k+k'-k''-1/2}\nn
d'^{(1),-A}_{-m}&=&\sum_{k=0}^{m-1}\sum_{k'=0}^{m-k-1}\sum_{k''\geq0}{}^{m-1}C_{k}{}^{-m}C_{k'}{}^{-m+k+k'}C_{k''}\nn
&&\quad{}\times(-1)^{m-k-1}a^{m+k''-k-1}(b-a)^{-m-k'}\tilde{d}^{-A,t\to\infty}_{-m+k+k'-k''+1/2}\cr
d'^{(2),+A}_{-m}&=&\sum_{k=0}^{m}\sum_{k'=0}^{m-k}\sum_{k''\geq0}\left({}^{m+2}C_{k}b-{}^{m}C_{k}a\right){}^{-m-1}C_{k'}{}^{-m+k+k'-1}C_{k''}\nn
&&\quad{}\times(-1)^{m-k}b^{m-k+k''+1}(a-b)^{-m-k'-1}\tilde{d}^{+A,t\to\infty}_{-m+k+k'-k''-1/2}\nn
d'^{(2),-A}_{-m}&=&\sum_{k=0}^{m-1}\sum_{k'=0}^{m-k-1}\sum_{k''\geq0}{}^{m-1}C_{k}{}^{-m}C_{k'}{}^{-m+k+k'}C_{k''}\nn
&&\quad{}\times(-1)^{m-k-1}b^{m+k''-k-1}(a-b)^{-m-k'}\tilde{d}^{-A,t\to\infty}_{-m+k+k'-k''+1/2}.
\label{InitialModes t=infinity}
\eea

\subsubsection*{\underline{$t=0$}}
\bea
d'^{(1),+A}_{-m}&=&-\sum_{k=0}^{m}\sum_{k'=0}^{m-k}\sum_{k''\geq0}{}^{-m-1}C_{k'}{}^{-m+k+k'-1}C_{k''}\left({}^{m+2}C_{k}a-{}^{m}C_{k}b\right)\nn
&&\quad{}\times(-1)^{m-k}a^{k'-k''}(b-a)^{-m-k'-1}\tilde{d}^{+A,t\to0}_{k''+1/2}\nn\nn
d'^{(1),-A}_{-m}&=&-\sum_{k=0}^{m-1}\sum_{k'=0}^{m-k-1}\sum_{k''\geq0}{}^{m-1}C_{k}{}^{-m}C_{k'}{}^{-m+k+k'}C_{k''}\nn
&&\quad{}\times(-1)^{m-k-1}a^{k'-k''-1}(b-a)^{-m-k'}\tilde{d}^{-A,t\to0}_{k''+1/2}\nn
d'^{(2),+A}_{-m}&=&-\sum_{k=0}^{m}\sum_{k'=0}^{m-k}\sum_{k''\geq0}{}^{-m-1}C_{k'}{}^{-m+k+k'-1}C_{k''}\left({}^{m+2}C_{k}b-{}^{m}C_{k}a\right)\nn
&&\quad{}\times(-1)^{m-k}b^{k'-k''}(a-b)^{-m-k'-1}\tilde{d}^{+A,t\to0}_{k''+1/2}\nn\nn
d'^{(2),-A}_{-m}&=&-\sum_{k=0}^{m-1}\sum_{k'=0}^{m-k-1}\sum_{k''\geq0}{}^{m-1}C_{k}{}^{-m}C_{k'}{}^{-m+k+k'}C_{k''}\nn
&&\quad{}\times(-1)^{m-k-1}b^{k'-k''-1}(a-b)^{-m-k'}\tilde{d}^{+A}_{k''+1/2}.
\label{InitialModes t=0}
\eea

We also present the zero modes of (\ref{InitialModes t=infinity}) and (\ref{InitialModes t=0}) explicitly.

\subsubsection*{Zero Modes at $t=\infty$}
\bea
d'^{(1),+A}_{0}&=& -\sum_{k''\geq0}(-1)^{k''}a^{k''}\tilde{d}^{+A,t\to\infty}_{-k''-1/2}\nn
d'^{(1),-A}_{0}&=&0\nn
d'^{(2),+A}_{0}&=&-\sum_{k''\geq0}(-1)^{k''}b^{k''+1}\tilde{d}^{+A,t\to\infty}_{-k''-1/2}\nn
d'^{(2),-A}_{0}&=&0.
\label{InitialZeroModes t=infinity}
\eea
\subsubsection*{Zero Modes at $t=0$}
\bea
d'^{(1),+A}_{-m}&=&\sum_{k''\geq0}(-1)^{k''}a^{-k''}\tilde{d}^{+A,t\to0}_{k''+1/2}\nn
d'^{(1),-A}_{-m}&=&0\nn
d'^{(2),+A}_{-m}&=&\sum_{k''\geq0}(-1)^{k''} b^{-k''}\tilde{d}^{+A,t\to0}_{k''+1/2}\nn
d'^{(2),-A}_{-m}&=&0.
\label{InitialZeroModes t=0}
\eea
As expected, any expression that tries to start with a negative zero mode vanishes.

\subsection{Computing the transition amplitudes}
Here we compute all of the various $f^{F\pm,(i)(j)+}_{mn}$ factors that include at least one zero mode.  We will do this by applying the various expressions for the transformed modes to Equation (\ref{tPlaneRelations}).  We repeat the relevant parts of this equation here for convenience.
\bea
f^{F-,(i)(1')}_{mn} &=& -{\nstbra d'^{(1'),++}_nd'^{(i),--}_{-m}\nstket\over\nstbra 0_{NS}\rangle_t}\nn
f^{F-,(i)(2')}_{mn} &=& -{\nstbra d'^{(i),--}_{-m}d'^{(2'),++}_n\nstket\over\nstbra 0_{NS}\rangle_t} \nn
f^{F+,(i)(1')}_{mn}&=&-{\nstbra d'^{(1'),--}_nd'^{(i),++}_{-m}\nstket\over\nstbra 0_{NS}\rangle_t\rangle}\nn
f^{F+,(i)(2')}_{mn}&=&-{\nstbra d'^{(i),++}_{-m}d'^{(2'),--}_n\nstket\over\nstbra 0_{NS}\rangle_t\rangle}.\label{FermionRelation}
\eea

We will now work through the different copy and charge combinations in turn.

\subsubsection*{Positive SU(2) R charge, final copy 1}
We begin with $f^{F+,(1)(1')}_{mn}$.  Since our transition amplitude is defined specifically for modes that do not annihilate the vacuum on which $|\chi\rangle$ is built, there is no case $n=0$ for this term.  Indeed, if one tries to evaluate this transition amplitude for $n=0$ it is found to vanish.

For the case $m=0$, we insert Equations (\ref{tplanemodesNS}) and  (\ref{InitialZeroModes t=infinity}) into Equation (\ref{FermionRelation}).  This gives:
\bea
f^{F+,(1)(1')}_{0,n}&=&-\frac{{}_{t}\langle 0_{NS}|d'^{(1'),--}_{n}d'^{(1),++}_{0}|0_{NS}\rangle_{t}}{{}_{t}\langle 0_{NS}|0_{NS}\rangle_{t}}\nn
&=&\sum_{j,j',k''\geq 0} {}^{n}C_j{}^{n}C_{j'}(-1)^{k''}a^{j+k''+1} b^{j'}\frac{{}_{t}\langle 0_{NS}|\tilde{d}^{--}_{n-j-j'-1/2}\tilde{d}^{++}_{-k''-1/2}|0_{NS}\rangle_{t}}{{}_{t}\langle 0_{NS}|0_{NS}\rangle_{t}}\quad \label{f11+ first}
\eea
Using the anticommutation relations from Equation (\ref{fermioncommutation}) along with creation/annihilation constraints, we find the following summation limits on the $j,j'$ sums:
\bea
n-j-j'-k''-1=0&\to&k'' = n-j-j'-1\nn
n-j-j'-1/2>0&\to&j'<n-j-1/2\nn
j'>0&\to&j<n-1/2.
\eea 
We then find:
\bea
\nn
f^{F+,(1)(1')}_{0,n}&=&-\sum_{j = 0}^{n-1} \sum_{j' = 0}^{n-j-1} {}^{n}C_j{}^{n}C_{j'}(-1)^{n-j-j'-1} a^{n-j'} b^{j'}.
\eea

We can now determine $f^{F+,(2)(1')}_{0,n}$ by applying the interchange $a\leftrightarrow b$.  This gives:
\bea
f^{F+,(2)(1')}_{0,n} &=& -\sum_{j = 0}^{n-1} \sum_{j' = 0}^{n-j-1} {}^{n}C_j{}^{n}C_{j'}(-1)^{n-j-j'-1} b^{n-j'} a^{j'}.
\eea

\subsubsection*{Positive SU(2) R charge, final copy 2}
We now turn to $f^{F+,(2)(2')}_{mn}$. Here both $m$ and $n$ are allowed to be zero.  We'll start with $m=0$.  Using (\ref{tplanezeromodesNS}),  (\ref{InitialZeroModes t=0}) and (\ref{FermionRelation}), we have:
\bea
f^{F+,(2)(2')}_{0,n} &=& \frac{{}_{t}\langle 0_{NS}|d'^{(2),++}_{0}d'^{(2'),--}_{n}|0_{NS}\rangle_{t}}{{}_{t}\langle 0_{NS}|0_{NS}\rangle_{t}}\label{F(f)22+_1}\\
&=& -\sum_{j,j',k''\geq 0}{}^{n}C_j{}^{n}C_{j'}(-1)^{k''}a^{n-j} b^{n-j'}b^{-k''}\frac{{}_{t}\langle 0_{NS}|\tilde{d}^{++,t\to0}_{k''+1/2}\tilde{d}^{--,t\to0}_{j+j'-n-1/2}|0_{NS}\rangle_{t}}{{}_{t}\langle 0_{NS}|0_{NS}\rangle_{t}}.\nonumber
\eea
We now use the anti commutation relations (\ref{fermioncommutation}) along with creation/annihilation constraints to limit the $j,j'$ sums:
\bea
k''+j+j'-n=0&\to& k''=n-j-j'\nn
j+j'-n-1/2< 0&\to& j'\leq n-j\nn
j'\geq 0&\to& j\leq n.
\eea
Using these constraints, (\ref{F(f)22+_1}) becomes:
\bea
f^{F+,(2)(2')}_{0,n}&=&-\sum_{j = 0}^{n}\sum_{j' = 0}^{n-j}{}^{n}C_j{}^{n}C_{j'}(-1)^{n-j-j'}a^{n-j} b^{j}.\label{fp220n}
\eea

Let us next examine the case $n=0$.  From (\ref{FermionRelation}) along with (\ref{tplanezeromodesNS}) and (\ref{InitialModes t=0}), we have:
\bea
f^{F+,(2)(2')}_{m,0} &=& \frac{{}_{t}\langle 0_{NS}|d'^{(2),++}_{m}d'^{(2'),--}_{0}|0_{NS}\rangle_{t}}{{}_{t}\langle 0_{NS}|0_{NS}\rangle_{t}}\nn
&=& \sum_{k=0}^{m}\sum_{k'=0}^{m-k}\sum_{k''\geq0}{}^{-m-1}C_{k'}{}^{-m+k+k'-1}C_{k''}\left({}^{m+2}C_{k}b-{}^{m}C_{k}a\right)\nn
&&\times(-1)^{m-k}b^{k'-k''}(a-b)^{-m-k'-1} { {}_{t}\langle 0_{NS}| \tilde{d}^{++,t\to0}_{k''+1/2}\tilde{d}^{--,t\to0}_{-1/2}|0_{NS}\rangle_{t} \over {}_{t}\langle 0_{NS}|0_{NS}\rangle_{t}} .
\eea
The anti commutation relations (\ref{fermioncommutation}) now gives only $k''=0$.  Thus:
\bea
f^{F+,(2)(2')}_{m,0} &=&  -\sum_{k=0}^{m}\sum_{k'=0}^{m-k}{}^{-m-1}C_{k'}\left({}^{m+2}C_{k}b-{}^{m}C_{k}a\right)(-1)^{m-k}b^{k'}(a-b)^{-m-k'-1}.
\eea
Naturally this expression should agree with (\ref{fp220n}) for the case $m=n=0$.  Indeed, both expressions yield:
\bea
f^{F+,(2)(2')}_{0,0} &=& 1.
\eea

We now assess the case of the initial mode living on copy 1 by applying the $a\leftrightarrow b$ interchange.  This gives:
\bea
f^{F+,(1)(2')}_{0,n} &=& -\sum_{j = 0}^{n}\sum_{j' = 0}^{n-j}{}^{n}C_j{}^{n}C_{j'}(-1)^{n-j-j'}b^{n-j} a^{j}\\
f^{F+,(1)(2')}_{m,0} &=&-\sum_{k=0}^{m}\sum_{k'=0}^{m-k}{}^{-m-1}C_{k'}\left({}^{m+2}C_{k}a-{}^{m}C_{k}b\right)(-1)^{m-k}a^{k'}(b-a)^{-m-k'-1}.\nonumber
\eea

\subsubsection*{Negative SU(2) R charge}
Here we turn to the amplitudes $f^{F-,(i)(j)}_{mn}$.  This amplitude vanishes for $m=0$, since an initial negative zero mode on either copy annihilates the R vacuum upon which our in state is built.  Similarly, a negative zero mode on copy 2 above the twists annihilates the vacuum upon which the out state is built.  We are thus left with only two amplitudes to consider, $f^{F-,(1)(1')}_{m,0}$ and $f^{F-,(2)(1')}_{m,0}$.  Since these are related by the $a \leftrightarrow b$ interchange, we will explicitly calculate only the first amplitude.
\bea
f^{F-,(1)(1')}_{m,0} &=& -\frac{{}_{t}\langle 0_{NS}|d'^{(1'),++}_{0}d'^{(1),--}_{-m}|0_{NS}\rangle_{t}}{ {}_{t}\langle 0_{NS}|0_{NS}\rangle_{t}}.\label{fminusratio}
\eea
Let us now examine (\ref{tplanezeromodesNS}).  The relevant transformed mode contains two parts:
\bea
d'^{(1'),+A}_{0}&=&\sum_{j,j'\geq 0}(-1)^{j+j'} a^j b^{j'} \tilde{d}^{+A,t\to\infty}_{-j-j'+1/2}-\sum_{j,j'\geq 0} (-1)^{j+j'} a^{j+1} b^{j'+1}\tilde{d}_{-j-j'-3/2}^{+A,t\to\infty}.
\eea
This is the leftmost mode, so we can drop any terms that annihilate $\nstbra$ on the left.  This includes the entirety of the second sum, as well as all terms from the first sum except $j=j'=0$.  Combined (\ref{InitialModes t=infinity}), Equation (\ref{fminusratio}) becomes:
\bea
f^{F-,(1)(1')}_{m,0} &=&-\sum_{k''\geq 0}\sum_{k=0}^{m-1}\sum_{k'=0}^{m-k-1}{}^{m-1}C_{k}{}^{-m}C_{k'}{}^{-m+k+k'}C_{k''}(-1)^{m-k-1}\\
&&\quad{}\times\left(a^{m+k''-k-1}(b-a)^{-m-k'}\frac{{}_{t}\langle 0_{NS}|\tilde{d}^{++,t\to\infty}_{1/2}\tilde{d}^{--,t\to\infty}_{-m+k+k'-k''+1/2}|0_{NS}\rangle_{t}}{ {}_{t}\langle 0_{NS}|0_{NS}\rangle_{t}}\right).\nonumber
\eea

We once again use the anticommutation relations (\ref{fermioncommutation}) to find the following constraints:
\bea
-m+k+k'-k''+1=0&\to&k''=-m+k+k'+1 \cr
 k''\geq 0&\to&k'\geq m-k-1
\eea
The $k'$ minimum here is already the maximum value it can take, so we can eliminate that sum in addition to eliminating the $k''$ sum.  We then find:
\bea
f^{F-,(1)(1')}_{m,0}&=&\sum_{k=0}^{m-1}{}^{m-1}C_{k}{}^{-m}C_{m-k-1}(-1)^{m-k-1}a^{m-k-1}(b-a)^{-2m+k+1}.
\eea
We now apply $a\leftrightarrow b$ to find:
\bea
f^{F-,(2)(1')}_{m,0}&=&\sum_{k=0}^{m-1}{}^{m-1}C_{k}{}^{-m}C_{m-k-1}(-1)^{m-k-1}b^{m-k-1}(a-b)^{-2m+k+1}.
\eea

\subsection{Summary of Results}
Here we gather all of the results for the fermionic transition amplitudes that involve at least one zero mode.
\bea
f^{F+,(1)(1')}_{0,n} &=&-\sum_{j = 0}^{n-1} \sum_{j' = 0}^{n-j-1} {}^{n}C_j{}^{n}C_{j'}(-1)^{n-j-j'-1} a^{n-j'} b^{j'}\nn
f^{F-,(1)(1')}_{m,0} &=&\sum_{k=0}^{m-1}{}^{m-1}C_{k}{}^{-m}C_{m-k-1}(-1)^{m-k-1}a^{m-k-1}(b-a)^{-2m+k+1}\nn
f^{F+,(2)(1')}_{0,n} &=& -\sum_{j = 0}^{n-1} \sum_{j' = 0}^{n-j-1} {}^{n}C_j{}^{n}C_{j'}(-1)^{n-j-j'-1} b^{n-j'} a^{j'}\nn
f^{F-,(2)(1')}_{m,0} &=&\sum_{k=0}^{m-1}{}^{m-1}C_{k}{}^{-m}C_{m-k-1}(-1)^{m-k-1}b^{m-k-1}(a-b)^{-2m+k+1}\nn
f^{F+,(2)(2')}_{m,0}&=& -\sum_{k=0}^{m}\sum_{k'=0}^{m-k}{}^{-m-1}C_{k'}\left({}^{m+2}C_{k}b-{}^{m}C_{k}a\right)(-1)^{m-k}b^{k'}(a-b)^{-m-k'-1}\cr
f^{F+,(2)(2')}_{0,n}&=&-\sum_{j = 0}^{n}\sum_{j' = 0}^{n-j}{}^{n}C_j{}^{n}C_{j'}(-1)^{n-j-j'}a^{n-j} b^{j}\nn
f^{F+,(1)(2')}_{m,0}&=&-\sum_{k=0}^{m}\sum_{k'=0}^{m-k}{}^{-m-1}C_{k'}\left({}^{m+2}C_{k}a-{}^{m}C_{k}b\right)(-1)^{m-k}a^{k'}(b-a)^{-m-k'-1}\nn
f^{F+,(1)(2')}_{0,n} &=& -\sum_{j = 0}^{n}\sum_{j' = 0}^{n-j}{}^{n}C_j{}^{n}C_{j'}(-1)^{n-j-j'}b^{n-j} a^{j}.
\eea

\section{Continuum Limit}\label{ContinuumLimit}

We have expressed the results of our computations as a set of finite sums over Binomial coefficients. These sums can be reduced to a single sum over a product of Hypergeometric functions, as was done in \cite{chm1}.  This expression using Hypergeometrics  is not particularly helpful, since the Hypergeometric functions are in general given by infinite series, and the truncation to a finite polynomial that actually happens here is not evident from their functional form. What we will find more useful is to consider the sums we have obtained for large mode numbers $n$. This is relevant for the actual physical problem that we wish to address, since the `effective string' that we have there will be very long, since it will be in a sector with high twist order. Then the excitations on this string occur in high mode numbers, and so form a continuous band rather than discrete levels. We call this limit of large mode numbers a continuum limit; this was studied for the single twist case in \cite{cmt}.

  To relate our computations so far to our physical problem, the first thing we need  to do is Wick rotate the time direction of the CFT to  Lorentzian signature. We will also make a simple choice of origin for the coordinates on the cylinder, placing our two twist operators symmetrically around this origin. This will lead to a simplifications of our expressions. We will then plot the finite sums we have obtained using a computer, and extract simple approximate forms for these sums from the plots. These expressions then encode the physics that we are trying to extract from the deformation of the CFT to second order. 
  
\subsection{Helpful Coordinates}
Let us first rotate the Euclidean time coordinate back to Minkowski time.
\bea
\t \to it &\implies& w = i(t+\s).
\eea
Now $w$ is purely imaginary.  At this point we are free to choose an origin for our coordinate system.  The most convenient is a point midway between $w_1$ and $w_2$, such that:
\bea
t_2 = -t_1 = {\D t\over2}, &\quad& \s_2 = -\s_1 = {\D\s\over2}.
\eea
This is the same choice of coordinate found in \cite{chm1}.  With our choice of branch cuts, this gives:
\bea
a &=& \cos^2\left({\D w \over 4i}\right)\nn
b &=& -\sin^2\left({\D w \over 4i}\right) ~=~ a-1.
\eea
As expected we now have a one-parameter system.  Our only relevant spacetime coordinate is now the fully-imaginary twist separation $\D w$.  The physics is also periodic in this coordinate, with a period of $4\pi i$.

\subsection{Numerical Analysis}
We now turn to the task of finding a good approximation for our transition amplitudes when the mode numbers become large.  We do this by plotting the exact values over a range of $\D w$ coordinates and then attempting to fit the resulting points.  We find a good, simple approximation for the bosonic transition amplitudes, which are themselves related to the fermionic transition amplitudes for nonzero modes.  The fermion zero mode functions prove less amenable to this analysis.  This might be expected, as for such amplitudes at least one mode is necessarily well outside of the continuum limit.

\begin{figure}[bht]
\includegraphics[width=0.5\columnwidth]{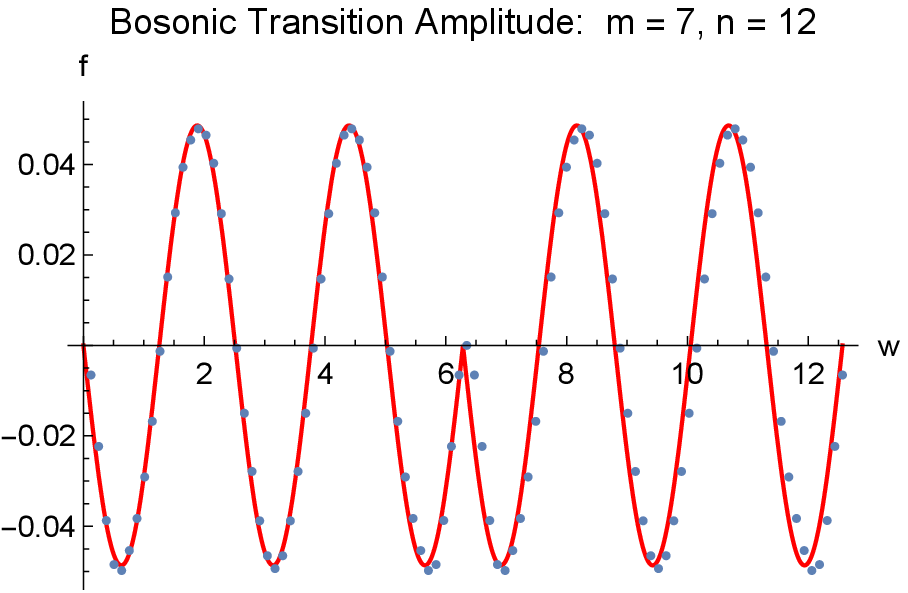} \includegraphics[width=0.5\columnwidth]{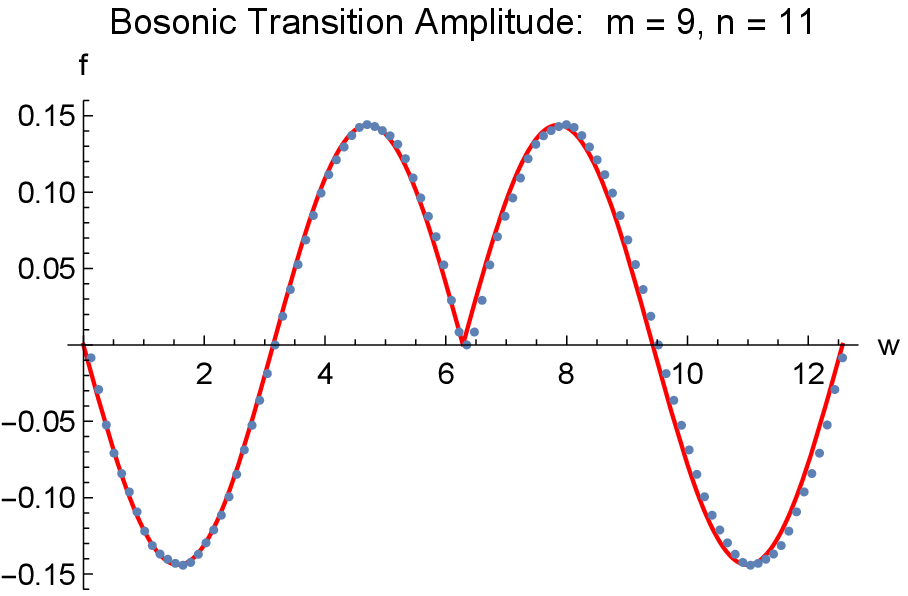}\\\\\\
\includegraphics[width=0.5\columnwidth]{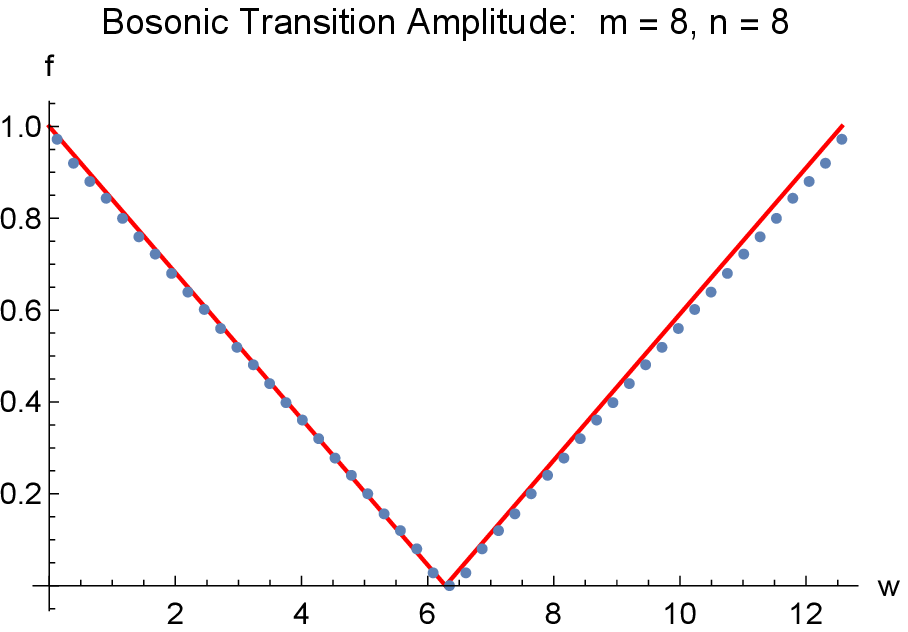} \includegraphics[width=0.5\columnwidth]{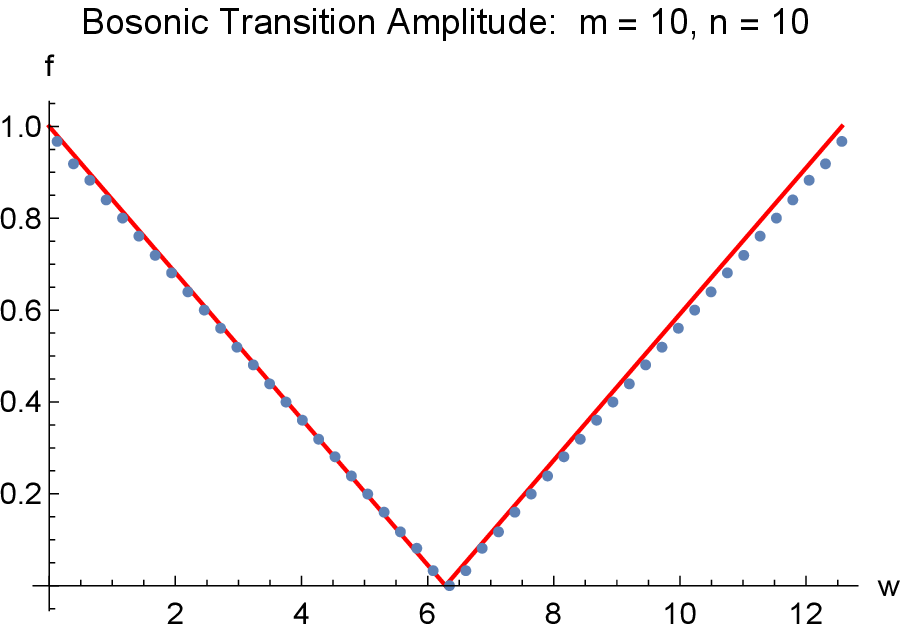}
\caption{The bosonic transition amplitudes $f^{B,(1)(1')}_{mn}$ for: $m=7$, $n=12$ (upper left); $m=9$, $n=11$ (upper right); $m=n=8$ (lower left); $m=n=10$ (lower right).  The exact values are plotted as blue dots, while the simple continuum limit approximation from Equation (\ref{CLApprox}) is plotted as a red line.  The axis labeled w is the real value $w/i$.}
\label{BosonGraphs}
\end{figure}

For the bosonic transition amplitude, large indices result in the approximate form (i.e., in the limit $m, n \gg 1$):
\bea
f^{B,(1)(1')}_{mn} &\approx& \begin{cases}
{1\over (m-n)\pi} \sqrt{m\over n}\sin\left((m-n){\D w\over 2i}\right)\text{sgn}\left({\D w \over 2\pi i}-1\right) & m\neq n\\
\left|{\D w \over 2\pi i}-1\right| & m=n.
\end{cases}\label{CLApprox}
\eea
This produces a mirroring about the point $\D w = 2\pi i$:
\bea
\left[f^{B,(1)(1')}_{mn}\right]_{\D w = 2\pi i + x} &=& \left[f^{B,(1)(1')}_{mn}\right]_{\D w = 2\pi i - x}.
\eea

Looking numerically at modes on the initial copy 2, we find a simple relationship:
\bea
f^{B,(1)(1')}_{mn} + f^{B,(2)(1')}_{mn} &=& \d_{mn}.\label{DeltaRelation}
\eea
While we have not proven this relationship in full generality, it has held for all cases we have checked including small $m$ and $n$.  There is also an analogous exact relationship at first-order:
\bea
f^{B,(1)}_{mn} + f^{B,(2)}_{mn} &=& \d_{mn}.
\eea
We thus expect (\ref{DeltaRelation}) to be an exact one.  Combined with the symmetry under global copy redefinitions:
\bea
f^{B,(1)(1')}_{mn} &=& f^{B,(2)(2')}_{mn}\nn
f^{B,(2)(1')}_{mn} &=& f^{B,(1)(2')}_{mn},
\eea
we find a simple interpretation of this symmetry: an excitation in the initial sate that is symmetric between the two copies will be left unaffected by the action of the two twists that we apply.  Schematically,
\bea
\sh\left(\a^{(1)} + \a^{(2)}\right)|\emptyset\rangle &=& \left(\a^{(1)} + \a^{(2)}\right)\sh|\emptyset\rangle.
\eea
A similar result holds for nonzero fermion modes.

Combining these results with the relationships found in Section \ref{Relations}, we see that there is only a single linearly-independent transition amplitude for both bosons and fermions with nonzero mode number.  A similar such simplification was also found for the $\g$ coefficients for the $|\chi\rangle$ state in  \cite{chm1}.

\subsection{Continuum Limit Conjecture}
Here we  note another a simple pattern among the transition amplitudes.  Consider the single-twist transition amplitudes found in \cite{acm2}. To relate these to the amplitude we have considered above for the 2-twist case, we make a simple change of notation: the mode numbers $p$ in \cite{acm2} need to be altered as   $p\to p/2$; with this change the mode numbers directly reflect the energy of the mode, which is the notation that we adopted in the present paper.    We then take the continuum limit, which corresponds to large mode numbers.  We use the approximations valid for large $n$:
\bea
{\G\left(n+\h\right)\over \G(n)} &\approx& \sqrt{n}.
\eea
From this, we find for the single twist case:
\bea
f^{B,(1)}_{np} &\approx& \begin{cases}
{1\over (n-p)\pi}\sqrt{n\over p} & n\neq p\\
\h & n=p
\end{cases}\label{OneTwistTA}
\eea

Consider the generic case $n\neq p$ in (\ref{OneTwistTA}). Compare this to result in the 2-twist case:
\bea
f^{B,(1)(1')}_{mn} &\approx& {1\over (m-n)\pi}\sqrt{m\over n} \sin\left((m-n){\D w \over 2\pi i}\right)\text{sgn}\left({\D w \over 2\pi i}-1\right).
\eea
We see that the 2-twist result has a structure similar to the single twist cae, apart from an extra oscillating factor with amplitude unity that appears becuase we now have a new parameter: the distance between the two twists.  A similar pattern is found in the Bogoliubov coefficients $\g$ from the state $|\chi\rangle$.  Using  \cite{chm1} and  \cite{acm1,acm2}, and again setting $w_0 = 0$, we have:
\bea
\g^B_{mn} &\approx& {1\over \sqrt{mn}(m+n)}\nn
\g^{B,(1')(1')}_{mn} &\approx& {1\over \sqrt{mn}(m+n)}h(m,n,\D w),
\eea
Here the function $h$ is again an oscillating factor, but we were unable to find a useful analytic approximation for this oscillating factor.  These observations suggest a  conjecture:

\subsubsection*{Conjecture}
When expressed in terms of excitation energies, the continuum limit forms of both the transition amplitudes and the Bogoliubov coefficients have the same amplitude at all orders of the twist operator; the only extra factors are oscillating functions that encode the separations between the twists.\\

In other words, we suspect that the patterns found between the first and second order calculations performed thus far may hold to arbitrary orders.

\section{What is the signal of thermalization?}

Our final goal is to understand  black hole formation in the CFT description. It is generally agreed that in this description, the formation of the hole corresponds to a process of thermalization. There have been several other approaches to thermalization and its relation to black hole formation; in this section we compare our approach with some of these other works. 

\subsection{Quantum quenches}

 Some effort has been directed to the evolution of a CFT state after a quantum quench \cite{cardy,mandal}. In this approach one starts with a specific type of initial state, and this evolves to a state with thermal properties. But one can perform this analysis even for a free CFT, where we do not expect thermalization. This suggests that the thermal behavior we are seeing in quantum quenches is somewhat different from the behavior we are looking for in the case of black hole formation. 
 
 In the black hole problem, we wish to start from a low entropy state --  perhaps one consisting of just two high energy particles -- and see this simple state evolve to a state with a  large number of low energy excitations. More precisely, if we look at the emission from this  latter state (following the  methods of \cite{radiation1}), then we should get an emission of many low energy quanta rather than the few high energy quanta that we started with. If we took a free theory, then there would be no such evolution; if we put high energy quanta in, then the emission will also correspond to high energy  quanta. 
 
 In our approach, we have studied the interacting D1D5 CFT by starting with the free theory and adding interactions to second order. Thus our goal is to find the `interaction vertex' which describes how a high energy excitation splits into several low energy excitations because of the deformation operator taking us away from the free theory. This interaction vertex is analogous to the $2 \rightarrow 2$ scattering of atoms in a gas; once we know how this scattering redistributes momenta after the collision, we can get a picture of how a gas starting in a non-thermal state relaxes to the thermal state after a large number of such collisions.

 \subsection{Redshift vs thermalization}
 
 We note that there are two different effects in the gravity theory, and they should have two different representations in the CFT:
 
 \bigskip
 
 (i) The first effect is the notion of gravitational redshift. A clock placed in the gravitational field of a mass $M$ slows down; this slowdown is given  by the local value of  $(g_{tt})^{\h}$. This redshift is produced by every body; it is not particular to the black hole. Of course the black hole is characterized by a redshift which {\it diverges} at the horizon (i.e., $g^{tt}\rightarrow \infty$), and the presence of such a diverging redshift is an indicator that a black hole  has formed or is about to form. But as we will argue below, the occurrence of a large redshift is not itself the process of thermalization. 
  
 \bigskip
 
 (ii) The second effect is thermalization. This may be simpler to define in the CFT rather than the gravity theory. The state of the field theory tends towards a typical or generic one for the given total energy, and it differs from the simple state which would be expected to define the initial high energy particles that formed the black hole. But what would be the gravity dual of such thermalization? Each state of the CFT should correspond to a state in the gravity theory. Further, it appears that the time coordinate $t$ of the CFT describes only the time coordinate outside the black hole horizon; thus if we have a collapsing shell, then the CFT description can follow this shell up to the point $r=r_0 +\e$ where $r_0$ is the radius of the horizon and $\e \ll 1$.  Thus if the CFT state changes to a generic state, then we should see a corresponding change in the gravity description {\it before} the shell reaches the horizon. 
 
 Such a change in the gravity description is possible only in the fuzzball paradigm \cite{fuzzball}; here the state of a collapsing shell tunnels into a linear combination of fuzzball states as the shell approaches $r\rightarrow r_0 +\e$. If the shell passes smoothly through its horizon in our gravity theory, we will see no change in its gravity state near $r=r_0$. In this case there will be no change in the CFT state from a special (low entropy) state to a generic (high entropy) state.
 
 \bigskip
 
 Let us now see the difference between (i) and (ii). When a body produces redshift, it does not have to change its own internal coherence; for example, the earth produces redshift without destroying its own atomic structure. Thus one can have redshift without thermalization. But if black hole formation is to be described by a thermalization in the CFT, then this thermalization has to correspond to a destruction of the  collapsing object as $r\rightarrow r_0 +\e $. Finding the physical effects that lead to such a destruction is of course the basic question behind the information paradox, and is resolved in the fuzzball paradigm through the discovery of fuzzball microstates and the idea of `entropy enhanced tunneling' into these microstates as the shell approaches its horizon. 
 
 \subsection{Finding redshift in gravity}
 
 In a very nice set of papers, the existence of gravitational redshift was shown in the CFT \cite{kaplan,kaplan2}. The basic idea was to look at the identity conformal block; this corresponds to exchange of the stress tensor  in the CFT, which corresponds to exchanging a graviton in the 
 gravity description. The whole Virasoro block then describes the exchange of an arbitrary number of gravitons. Using this technique, it was shown that excitations in the background of a massive particle had the correct redshift; in fact a direct link was made between the  CFT energies and the conical defect metric in the dual gravitational theory. In \cite{mathur1} the notion of gravitational redshift was related directly to the existence of an energy gap in the CFT. In the presence of such a gap, excitations of some degrees of freedom imply an effective stress tensor acting on other degrees of freedom, and the Ward identity then relates this stress tensor to a slowdown of effective time. There is a nice description of such Virasoro blocks on the gravity side, discussed for example in \cite{kraus}.
 
 In \cite{hartman} an explicit model of shell collapse was studied, using the methods of \cite{kaplan}. The shell was represented by a number of infalling particles in the gravity description, and a corresponding set of operators in the dual CFT. In the background of this shell, the 2-point function of a light operator was computed. When the time separation between the points in this correlator was taken to be 
 large, it was found that the correlator became exponentially small. This was taken as evidence of black hole formation; since the exponential decay matched  a similar decay computed in \cite{esko} in the gravity description.
 
Let us therefore take a more detailed look at the gravity computation of \cite{esko}, and thereby deduce what the exponential decay found there should signify in the CFT. The gravity computation started with a shell that moved inwards in AdS space. The motion of this shell was followed up to the horizon, but not past the horizon. A light field $\phi$ was considered in the background produced by this collapsing shell. The correlator $\langle\phi(t_2)\phi(t_1)\rangle$ was computed for two points on the boundary of AdS. 

To compute this correlator, one solves the wave equation for $\phi$ in the background of the collapsing shell. If $t_2-t_1$ is large, then the insertion $\phi(t_1)$ produces a waveform which falls in from the boundary to the location of the collapsing shell, which in turn is just outside its own horizon $r_0 + \e$. A small part of this  waveform then leaks out again to infinity, and this part is responsible for producing the correlator $\langle\phi(t_2)\phi(t_1)\rangle$. Thus the smallness of the correlator can be seen to arise from the difficulty of emerging from a region of large redshift; in other words, the exponential decay of this correlator is a consequence of large redshift.

 But in this gravity picture, {\it the shell has not disintegrated in any way}; i.e., it still has its original internal structure, and has not for example become a linear combination of fuzzballs. Thus by what was noted above, the dual CFT state should also be a state where the shell state has not undergone a change to a generic CFT configuration. Thus we are not seeing thermalization; rather, we are seeing the effects of large redshift.
 
   \subsection{Seeing thermalization}
   
   Let us now ask: what should thermalization look like in the CFT? We should see a spread into a new set of states, which describe a `quark gluon plasma' rather than a few simple supergravity particles. These generic plasma states have a energy and  entropy that is governed by $c$; thus these are a large set of new states that we expect to arise when we have sufficient energy to create the plasma. 
   
   Thus we expect the following. If we have states below the threshold of black hole formation, then they remain essentially unchanged in the evolution which   in the dual gravity description corresponds to `falling inwards to smaller $r$'. But if the energy of the state is above the black hole threshold, then at some radius $r=r_0 +\epsilon$, the state undergoes a significant change of character. In the CFT this change of character is `thermalization', while in the gravity description this is the `transition to fuzzballs'.\footnote{  Recently an interesting work \cite{giusto} emphasized the role of conformal blocks other than the identity in maintaining the unitarity of correlators in the CFT. }
   
   \subsection{The role of large entropy}
   
   To make the above expectation more concrete, suppose the shell in gravity is made of $n$ high energy  gravitons. When the shell is far from its horizon, we should be able to identify these separate gravitons in the CFT state; each graviton should correspond to a singlet `hadron' in the gauge theory. As long as we can identify the $n$ separate particles in the CFT state, we should say that the state has not thermalized. 
   
   In the fuzzball paradigm, what happens as $r\rightarrow r_0 + \e$ is the following. There is an amplitude $A$ to tunnel from the shell state to a fuzzball state. This amplitude is very small, since it corresponds to tunneling between two macroscopic gravity solutions but the number of possible fuzzball solutions that  we can tunnel to is very large, given by $Exp[S_{bek}]$. We get
   \be
   |A|^2 Exp[S_{bek}] \sim 1
   \ee
   so that the collapsing shell tunnels into fuzzballs and we do not get the traditional black hole \cite{tunnel}. 
   
   Now let us ask what this physics corresponds to in the dual CFT. Each fuzzball state should be dual to a state with high dimension $\Delta$. While the transition from our initial state to this high $\Delta $ state has a small probability, the number of these high dimension states is large ($\sim Exp[S_{bek}]$). The net effect is that the state corresponding to $n$ gravitons transitions to a very different state which is a superposition of these dimension $\sim \Delta$ states. This is the transition to the quark-gluon plasma in the CFT, and is the true indicator of thermalization. 
   
   Now we can see the difficulty we are pointing to. In the gravity computation of \cite{esko} mentioned above, the infalling shell did not break up into anything; it simply approached the horizon, maintaining its internal structure.  In fact for the purposes of the computation of \cite{esko}, we could take the classical picture where the  shell  passed smoothly into the interior of the horizon without any change to its internal structure. Thus we have no breakup into fuzzball states, and thus no map to states of a quark gluon plasma in the dual CFT. But even without the breakup of the shell, the two point function $\langle \phi(t_1)\phi(t_2)\rangle$ is found to decay exponentially with $t_2-t_1$. As mentioned above, this decay just follows from the fact that the wavefunction created by $\phi(t_1)$ falls to a location near the horizon, and then only an exponentially small part can make it back to the boundary at $t_2$. The exponential decay in this case is created by the large redshift of the shell near its horizon, and not by the breakup of the shell. Thus we see that while exponential falloffs in correlators can be caused by thermalization, they can also be caused by large redshift.

   \subsection{Comparing to our approach}

   How do we know if the exponential falloff is caused by thermalization or redshift? If we look at correlators with a small number of points in the CFT, then  it is possible that we will not be able to tell the difference; just as in the case of the 2- point function above, the dominant fall off will arise from the large redshift near the shell which is just outside the horizon. Thus the natural object to look at is the entire wavefunction of the CFT state (rather than low point correlators). This is of course difficult, and so we have had to follow a perturbative approach; we start with the free CFT (which has no thermalization) and then go to second order in the twist perturbation so that we extract the basic `scattering vertex' of the theory. As we hope to show in a following paper, an initial perturbation does show signs of thermalization due to  this second order vertex.
   
    How can we relate our approach to the abstract treatment of large $c$ CFTs? The most promising approach seems to be that in \cite{kaplan2}, where the identity block of the 4-point function was computed in a large $c$ approximation and continued to   {\it Lorentzian} signature. Two of the points were heavy operators, while two were light.  When the mass of the heavy operator exceeded the black hole bound, then it was found that the correlator exponentially decayed, showing that the light particle falling towards the heavy particle had a small chance of escaping away again. We can consider this as a t-channel process, where the stress tensor exchange generates a large redshift on the light particle due to the heavy particle. What would be useful to see would be an s-channel description, where we have two particles, each below the black hole threshold, collide and form an object above the threshold. In this channel one should then see the effect of entropy overwhelming the energy for formation of heavy operators, in line with the fuzzball paradigm on the gravity side. 
    
    \subsection{The effective horizon in the fuzzball paradigm}
    
    We comment on what one expects about the behavior of horizons in the fuzzball paradigm. The conjecture of fuzzball complementarity says that while there is no real horizon in a fuzzball state, interacting with this state `feels' like a horizon for objects that are infalling with $E\gg T$; i.e., with energy much larger than the temperature of the black hole. Thus the effective infall through the horizon is supposed to emerge in an `approximate dual description'.  In spirit the idea here is similar to the idea of AdS/CFT duality, where the dual map is {\it exact}. In AdS/CFT, there is a description of D-branes where an infalling graviton breaks up into open strings when hitting these branes. But the open strings produced in this process are created in a particular coherent state, and so one has an effective dual representation where the infalling graviton falls smoothly into an AdS region. In the black hole, the detailed state written in terms of fuzzballs is like the open string description, while the effective infall into the classical metric emerges as an approximate behavior (in the $E\gg T$ limit), and is analogous to the AdS infall. The need for the approximation $E\gg T$ is crucial, since we need to carry out the information of the state in $E\sim T$ quanta.  These $E\sim T$ quanta are not described by evolution in the traditional black hole semiclassical geometry, and this is what resolves the information paradox.

\section{Discussion}

Our goal is to see thermalization in the D1D5 CFT. This thermalization is expected to be dual to the process of black hole formation in the gravity theory. Thus if we can understand thermalization in the CFT, we get a window on the some of the deepest questions associated to black holes. 

The free D1D5 CFT itself has been very useful; it gives correctly the extremal and near-extremal entropies of the black hole, as well as the greybody factors of radiation from the near-extremal hole. But the free theory cannot show thermalization, since the excitations on the D1D5 brane system are free fields at the orbifold point. To reach the supergravity point, we have to consider the deformation operator $O$ that takes us away from the orbifold point.

The application of a single deformation $O_{\dot{A}}$ has been studied extensively in previous work. Single twist Bogoluibov coefficients $\g^B_{mn},\g^F_{mn}$, transition amplitudes $f^{B}_{np},f^{F\pm}_{np}$, as well as the application of the supercharge contours on both the twisted vacuum and an initial excitation were computed. The generalization of these quantities to arbitrary initial winding number were computed as well. However, no clear evidence for thermalization was seen at this order. Rather, we saw a vanishing contribution for the splitting of an initial wavepacket from the full deformation in the limit of large $m,n$. We therefore extended our analysis to the case of two deformations, $O_{\dot{A}}O_{\dot{B}}$. At this order removing the $G$ contours from the twists in order to compute the corresponding Bogoliubov coefficients proved considerably more difficult, but still manageable. In our previous paper we computed the two twist Bogoliubov coefficients, $\g^{B}_{mn},\g^{F\pm}_{mn}$. Interestingly, the two twist behavior at large values of $m,n$ seemed to match the single twist behavior up to a complicated oscillating factor, involving the twist separation $\Delta w$. Because this is a supersymmetric theory, we also found relationships between the two twist $\g^B_{mn}$ and $\g^{F}_{mn}$ for all nonzero modes. 

In this work we extend our previous analysis to include $1$-loop transition amplitudes where the initial state is not the vacuum. We start with two singly wound copies, one with an initial excitation, and after applying two twists $\s_2^+\s_2^+$ we return to two singly wound strings with a sum over creation operators weighted by functions $f^{B},f^{F\pm}$, which we compute in a similar manner as in one twist case. We note that the zero mode analysis for the two twist $f^{F\pm}$'s was an added complication absent from the single twist case.  Again, with the introduction of the twist separation, $\Delta w$, new and interesting behavior arises. We now find that an initial mode starting on Copy $1$ will create a distribution of final modes peaking around the initial energy on Copy $1$ for very small twist separation that then, for increasing $\Delta w$, transitions onto Copy $2$  and then back to Copy $1$ after a period of $4\pi$. This behavior is symmetric under interchange of Copy $1$ and Copy $2$. This is consistent with what we would expect as now the initial mode can travel back and forth between the two strings. For large values of $n,p$ we again find similar behavior with the single twist transition amplitudes differing by a simple oscillating factor, which we identified. Just as for the Bogoliubov coefficients, supersymmetry produces relations between $f^{B}$ and $f^{F\pm}$ as well. In the appendix we also computed the two twist boson wick contraction term and showed agreement with the $\g^{B}$ for the reverse process. We did not explicitly compute the fermion two twist wick contraction but noted that the procedure was similar to the bosonic computation with additional spectral flow complications. The wick term is actually related to $\g^{F}$ of a state, $|\chi'\rangle$, that was not computed in this or any previous work. However, it provides an easier method for computing the two twist fermion wick contraction term if necessary.

Having computed both the Bogoliubov coefficients and the transition amplitudes, in order to obtain a complete analysis, we still must apply the supercharge contours arising from the deformation operators. We hope to return to this in a future work.\\\\

\section*{Acknowledgements}
This work is supported in part by DOE grant de-sc0011726.\\\\
\appendix
\section{CFT notation and conventions} \label{ap:CFT-notation}

We follow the notation of \cite{acm1, acm2}, which we record here for convenience.
We have 4 real left moving fermions $\psi_1, \psi_2, \psi_3, \psi_4$ which we group into doublets $\psi^{\alpha A}$ as follows:
\be
\begin{pmatrix}
\psi^{++} \cr \psi^{-+}
\end{pmatrix}
=\sqi
\begin{pmatrix}
\psi_1+i\psi_2 \cr \psi_3+i\psi_4
\end{pmatrix}
\ee
\be
\begin{pmatrix}
\psi^{+-} \cr \psi^{--}
\end{pmatrix}
=\sqi
\begin{pmatrix}
\psi_3-i\psi_4 \cr -(\psi_1-i\psi_2)
\end{pmatrix}.
\ee
Here $\alpha=(+,-)$ is an index of the subgroup $SU(2)_L$ of rotations on $S^3$ and $A=(+,-)$ is an index of the subgroup $SU(2)_1$ from rotations in $T^4$. The reality conditions on the individual fermions are
\bea
(\psi_i)^{\dagger} = \psi_i \qquad \Rightarrow \qquad (\psi^{\a A})^{\dagger} = - \epsilon_{\alpha\beta}\epsilon_{AB} \psi^{\beta B} \,.
\eea 
One can introduce doublets $\psi^\dagger$, whose components are given by
\bea
(\psi^\dagger)_{\alpha A} &=& (\psi^{\a A})^{\dagger},
\eea 
from which the reality condition is given by
\be
 (\psi^\dagger)_{\alpha A}=-\epsilon_{\alpha\beta}\epsilon_{AB} \psi^{\beta B}.
\ee
The 2-point functions are
\be
<\psi^{\alpha A}(z)(\psi^\dagger)_{\beta B}(w)>=\delta^\alpha_\beta\delta^A_B{1\over z-w}, ~~~
<\psi^{\alpha A}(z)\psi^{\beta B}(w)>=-\epsilon^{\alpha\beta}\epsilon^{AB}{1\over z-w},
\ee
where we have:
\be
\epsilon_{12}=1, ~~~\epsilon^{12}=-1, ~~~
\psi_A=\epsilon_{AB}\psi^B, ~~~
\psi^A=\epsilon^{AB}\psi_B \,.
\ee
There are 4 real left moving bosons $X_1, X_2, X_3, X_4$, which can be grouped into a matrix:
\be
X_{A\dot A}= \sqi X_i \sigma_i
=\sqi
\begin{pmatrix}
X_3+iX_4 & X_1-iX_2 \\ X_1+iX_2&-X_3+iX_4
\end{pmatrix},
\ee
where $\sigma_i=(\sigma_a, iI)$. The reality condition on the individual bosons is given by
\bea
(X_i)^{\dagger} = X_i \qquad \Rightarrow \qquad (X_{A\dot A})^{\dagger} = - \e^{AB}\e^{\dot A \dot B} X_{B \dot B} \,.
\eea 
One can introduce a matrix, $X^\dagger$, with components  
\be
(X^\dagger)^{A\dot A}~=~ (X_{A\dot A})^{\dagger}~=~\sqi
\begin{pmatrix}
X_3-iX_4& X_1+iX_2\\
X_1-iX_2&-X_3-iX_4
\end{pmatrix},
\ee
from which the reality condition is given by
\bea
(X^\dagger)^{A\dot A}~=~ - \e^{AB}\e^{\dot A \dot B} X_{B \dot B} \,.
\eea 
The 2-point functions are
\be
<\partial X_{A\dot A}(z) (\partial X^\dagger)^{B\dot B}(w)>=-{1\over (z-w)^2}\delta^B_A\delta^{\dot B}_{\dot A}, ~~~
<\partial X_{A\dot A}(z) \partial X_{B\dot B}(w)>={1\over (z-w)^2}\epsilon_{AB}\epsilon_{\dot A\dot B} \,.
\ee

The chiral algebra is generated by the operators
\be
J^a=-{1\over 4}(\psi^\dagger)_{\alpha A} (\sigma^{Ta})^\alpha{}_\beta \psi^{\beta A}
\ee
\be
G^\alpha_{\dot A}= \psi^{\alpha A} \partial X_{A\dot A}, ~~~(G^\dagger)_{\alpha}^{\dot A}=(\psi^\dagger)_{\alpha A} \partial (X^\dagger)^{A\dot A}
\ee
\be
T=-{1\over 2} (\partial X^\dagger)^{A\dot A}\partial X_{A\dot A}-{1\over 2} (\psi^\dagger)_{\alpha A} \partial \psi^{\alpha A}
\ee
\be
(G^\dagger)_{\alpha}^{\dot A}=-\epsilon_{\alpha\beta} \epsilon^{\dot A\dot B}G^\beta_{\dot B}, ~~~~G^{\alpha}_{\dot A}=-\epsilon^{\alpha\beta} \epsilon_{\dot A\dot B}(G^\dagger)_\beta^{\dot B} \,.
\ee
These operators generate the OPE algebra
\be
J^a(z) J^b(z')\sim \delta^{ab} {\h\over (z-z')^2}+i\epsilon^{abc} {J^c\over z-z'}
\ee
\be
J^a(z) G^\alpha_{\dot A} (z')\sim {1\over (z-z')}\h (\sigma^{aT})^\alpha{}_\beta G^\beta_{\dot A}
\ee
\be
G^\alpha_{\dot A}(z) (G^\dagger)^{\dot B}_\beta(z')\sim -{2\over (z-z')^3}\delta^\alpha_\beta \delta^{\dot B}_{\dot A}- \delta^{\dot B}_{\dot A}  (\sigma^{Ta})^\alpha{}_\beta [{2J^a\over (z-z')^2}+{\partial J^a\over (z-z')}]
-{1\over (z-z')}\delta^\alpha_\beta \delta^{\dot B}_{\dot A}T
\ee
\be
T(z)T(z')\sim {3\over (z-z')^4}+{2T\over (z-z')^2}+{\partial T\over (z-z')}
\ee
\be
T(z) J^a(z')\sim {J^a\over (z-z')^2}+{\partial J^a\over (z-z')} 
\ee
\be
T(z) G^\alpha_{\dot A}(z')\sim {{3\over 2}G^\alpha_{\dot A}\over (z-z')^2}  + {\partial G^\alpha_{\dot A}\over (z-z')} \,.
\ee

Note that
\be
J^a(z) \psi^{\gamma C}(z')\sim {1\over 2} {1\over z-z'} (\sigma^{aT})^\gamma{}_\beta \psi^{\beta C} \,.
\ee

The above OPE algebra gives the commutation relations
\begin{eqnarray}
\com{J^a_m}{J^b_n} &=& \frac{m}{2}\delta^{ab}\delta_{m+n,0} + i{\epsilon^{ab}}_c J^c_{m+n}
            \\
\com{J^a_m}{G^\alpha_{\dot{A},n}} &=& \frac{1}{2}{(\sigma^{aT})^\alpha}_\beta G^\beta_{\dot{A},m+n}
             \\
\ac{G^\alpha_{\dot{A},m}}{G^\beta_{\dot{B},n}} &=& \hspace*{-4pt}\epsilon_{\dot{A}\dot{B}}\bigg[
   (m^2 - \frac{1}{4})\epsilon^{\alpha\beta}\delta_{m+n,0}
  + (m-n){(\sigma^{aT})^\alpha}_\gamma\epsilon^{\gamma\beta}J^a_{m+n}
  + \epsilon^{\alpha\beta} L_{m+n}\bigg]\quad\\
\com{L_m}{L_n} &=& \frac{m(m^2-\frac{1}{4})}{2}\delta_{m+n,0} + (m-n)L_{m+n}\\
\com{L_m}{J^a_n} &=& -n J^a_{m+n}\\
\com{L_m}{G^\alpha_{\dot{A},n}} &=& \left(\frac{m}{2}-n\right)G^\alpha_{\dot{A},m+n} \,.
\end{eqnarray}

\section{Ramond vacua notation}\label{RVN}
Here we define our notation for the various Ramond vacua in the untwisted sector.  There are two copies, which are not technically separate Hilbert spaces.  We start with the vacuum
\be
\rmutvket \equiv |v\rangle
\ee
and act on it with various fermion zero modes to construct the other Ramond vaccua.  In order to be consistent with \cite{acm1}, we also require something along the lines of
\be
\rptvket^{(i)} = d_0^{(i)++}d_0^{(i)+-}\rmtvket^{(i)},
\ee
though we do not actually have states containing only one of the two copies.

We now present a table defining our notation for the various vacua.
\bea
|v\rangle & = & \rmutvket \nn
d_0^{(1)+-}|v\rangle &=& |0_R\rangle^{(1)} \otimes |0_R^-\rangle^{(2)}\nn
d_0^{(1)++}|v\rangle &=& |\tilde{0}_R\rangle^{(1)} \otimes |0_R^-\rangle^{(2)}\nn
d_0^{(1)++}d_0^{(1)+-}|v\rangle &=& |0_R^+\rangle^{(1)} \otimes |0_R^-\rangle^{(2)}
\eea
\bea
d_0^{(2)+-}|v\rangle &=& |0_R^-\rangle^{(1)} \otimes |0_R\rangle^{(2)}\nn
d_0^{(1)+-}d_0^{(2)+-}|v\rangle &=& |0_R\rangle^{(1)} \otimes |0_R\rangle^{(2)}\nn
d_0^{(1)++}d_0^{(2)+-}|v\rangle &=& |\tilde{0}_R\rangle^{(1)} \otimes |0_R\rangle^{(2)}\nn
d_0^{(1)++}d_0^{(1)+-}d_0^{(2)+-}|v\rangle &=& |0_R^+\rangle^{(1)} \otimes |0_R\rangle^{(2)}
\eea
\bea
d_0^{(2)++}|v\rangle &=& |0_R^-\rangle^{(1)} \otimes |\tilde{0}_R\rangle^{(2)}\nn
d_0^{(1)+-}d_0^{(2)++}|v\rangle &=& |0_R\rangle^{(1)} \otimes |\tilde{0}_R\rangle^{(2)}\nn
d_0^{(1)++}d_0^{(2)++}|v\rangle &=& |\tilde{0}_R\rangle^{(1)} \otimes |\tilde{0}_R\rangle^{(2)}\nn
d_0^{(1)++}d_0^{(1)+-}d_0^{(2)++}|v\rangle &=& |0_R^+\rangle^{(1)} \otimes |\tilde{0}_R\rangle^{(2)}
\eea
\bea
d_0^{(2)++}d_0^{(2)+-}|v\rangle &=& |0_R^-\rangle^{(1)} \otimes |0_R^+\rangle^{(2)}\nn
d_0^{(1)+-}d_0^{(2)++}d_0^{(2)+-}|v\rangle &=& |0_R\rangle^{(1)} \otimes |0_R^+\rangle^{(2)}\nn
d_0^{(1)++}d_0^{(2)++}d_0^{(2)+-}|v\rangle &=& |\tilde{0}_R\rangle^{(1)} \otimes |0_R^+\rangle^{(2)}\nn
d_0^{(1)++}d_0^{(1)+-}d_0^{(2)++}d_0^{(2)+-}|v\rangle &=& |0_R^+\rangle^{(1)} \otimes |0_R^+\rangle^{(2)}.
\eea
These relations hold for both the initial (pre-twists) and final (post-twists) sectors.

%%%%%%%%%%%%%%%%%%%%%%%%%%%%%%%%%%%%%%%%%%%%%%%%%%%%%%%%55

\section{Computation of Two Twist Wick Contraction Term}
In this section we compute the two twist boson wick contraction term, $C^{B,2}_{++--}$.
\subsection{Computing $C^{B,2}_{++--}$} 
Before computing $C^{B,2}_{++--}$ let us first define some useful relations that we will need. Consider the wick contraction term for a single untwisting:
\bea
\nn
C^{B,1,\text{unt.}}_{mn,++--}&=&\langle 0|\s_{2}^{+}(w_0)\a_{++,-m}\a_{--,-n}|0\rangle\nn
&=&\left(\langle 0|\a_{++,n}\a_{--,m}\s_{2}^{+}(w_0)|0\rangle\right)^{\dagger}\nn
&=&-mn\left(\g^{B}_{mn}\right)^*\nn
\label{single untwisting wick contraction}
\eea 
where $\left(\g^{B}_{mn}\right)^{\dagger}=\left(\g^{B}_{mn}\right)^{*}$. Even though $\left(\s_{2}^{+}\right)^{\dagger}=\s_{2}^{-}$ we neglect the conjugation because the bosons are unaffected by spectral flow.
We see that the single untwisting wick contraction term is equal the conjugate of the single twist $\g^{B}$. There is one other relation we must consider before we tackle the two twist wick contraction term. Consider the single untwisting $f^{B}$ which we'll write as $f^{B,\text{unt.}}$:
 \bea
 \nn
 f^{B,\text{unt.}}_{pn} &=&{1\over n} \langle 0|\a^{(1)}_{++,n}\s_{2}^{+}(w_0)\a_{--,-p}|0\rangle \nn
 &=&{1\over n} \left(   \langle 0|\a_{++,p}\s_{2}^{+}(w_0)\a^{(1)}_{--,-n}|0\rangle  \right)^{\dagger}\nn
 &=&{p\over n}\left(f^{B}_{np}\right)^{*}\nn
 \label{twist untwist f relation}
 \eea
where $\left(f^{B}_{np}\right)^{\dagger} = \left(f^{B}_{np}\right)^{*}$. We see that the $f^{B,\text{unt.}}$ is just the conjugate transpose of the single twist $f^{B}$. Now that we have these two relations, let us compute the two twist wick contraction term. We begin with the following amplitude
%\bea
%\nn
%\mathcal{A}_1^{(i)(j)} &=& {}^{(1)}\langle 0 |  {}^{(2)}\langle 0 | \s_{2}^{+}(w_2)\s_{2}^{+}(w_1)\a_{++,-m}^{(i)}\a_{--,-n}^{(j)}|0\rangle^{(1)}|0\rangle^{(2)}\nn\nn
%&=& \left({}^{(1)}\langle 0 |  {}^{(2)}\langle 0 | \a_{++,n}^{(j)}\a_{--,m}^{(i)}\s_{2}^{+}(w_1)\s_{2}^{+}(w_2)|0\rangle^{(1)}|0\rangle^{(2)}\right)^*\nn\nn
%&=&(\g^{B(j)(i)}_{nm})^*
%\eea
%where $m,n>0$. We can also write the two twist wick contraction term in terms of single twist and untwist quantities. We again start of with our amplitude
\bea
\nn
C^{B,2}_{mn,++--} &=& {}^{(1)}\langle 0 |  {}^{(2)}\langle 0 | \s_{2}^{+}(w_2)\s_{2}^{+}(w_1)\a_{++,-m}^{(1)}\a_{--,-n}^{(1)}|0\rangle^{(1)}|0\rangle^{(2)}\nn
\eea
where $m,n>0$.
Now, bringing both $\a$'s through the first twist gives the following:
\bea
&&C^{B,2}_{++--,mn} \nn
&&\quad = {}^{(1)}\langle 0 |  {}^{(2)}\langle 0 | \s_{2}^{+}(w_2)\left(\sum_{p\geq 0}f_{mp}^{B}(w_1)\a_{++,-p}\right)\left(\sum_{p'}f_{np'}^{B}(w_1)\a_{--,-p'} \right)\s_{2}^{+}(w_1)|0\rangle^{(1)}|0\rangle^{(2)}\nn
&&\quad~~  + ~ C^{1}_{mn,++--}\nn
\eea
This step was computed in \cite{acm2}.
Let us now pull the two $\a$'s through the second twist. Doing this gives:
\bea
\nn
C^{B,2}_{mn,++--} &=& \sum_{p,p'\in \mathbb{Z}^{+}_{\text{odd}},q,q' > 0}f_{mp}^{B}(w_1)f_{np'}^{B}(w_1)f^{B,\text{unt.}}_{pq}(w_2)f^{B,\text{unt.}}_{p'q'}(w_2)\nn\nn
&&\quad\times {}^{(1)}\langle 0 | {}^{(2)}\langle 0 |\a_{++,-q}^{(1)}\a_{--,-q'}^{(1)}\s_{2}^{+}(w_2)\s_{2}^{+}(w_1)|0\rangle^{(1)}|0\rangle^{(2)}\nn\nn
&&\quad + \sum_{p,p' \in \mathbb{Z}^{+}_{\text{odd}} }f_{mp}^{B}(w_1)C^{B,1,\text{unt.}}_{pp',++--} f_{np'}^{B}(w_1) +  C^{1}_{mn,++--}\nn\nn
&=&\sum_{p,p' \in \mathbb{Z}^{+}_{\text{odd}} }f_{mp}^{B}(w_1)C^{B,1,\text{unt.}}_{pp',++--} f_{np'}^{B}(w_1) +  C^{1}_{mn,++--}\nn
\label{two twist wick contraction}
\eea
where again we use the result computed in \cite{acm2} and also the relation:
\bea
\nn
{}^{(1)}\langle 0 | {}^{(2)}\langle 0 |\a_{++,-q}^{(1)}\a_{--,-q'}^{(1)}\s_{2}^{+}(w_2)\s_{2}^{+}(w_1)|0\rangle^{(1)}|0\rangle^{(2)}=0\nn
\eea
since the two $\a$'s annihilate on the left.

Now using the relations found in (\ref{single untwisting wick contraction}) and (\ref{twist untwist f relation}), (\ref{two twist wick contraction}) becomes:
\bea
\nn
C^{B,2}_{mn,++--} &=& -\sum_{p,p' \in \mathbb{Z}^{+}_{\text{odd}}}pp'f_{mp}^{B}(w_1)\left(\g^{B}_{pp'}(w_2)\right)^* f_{np'}^{B}(w_1) +  C^{1}_{mn,++--}.\nn
\label{two twist wick final}
\eea
\begin{figure}
\includegraphics[width=0.5\columnwidth]{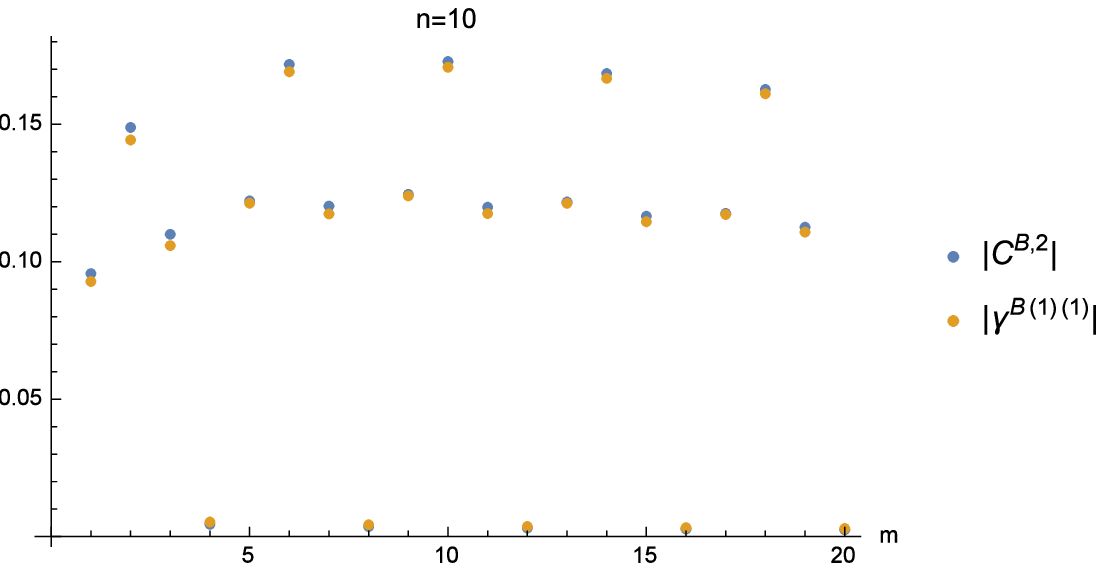}
\includegraphics[width=0.5\columnwidth]{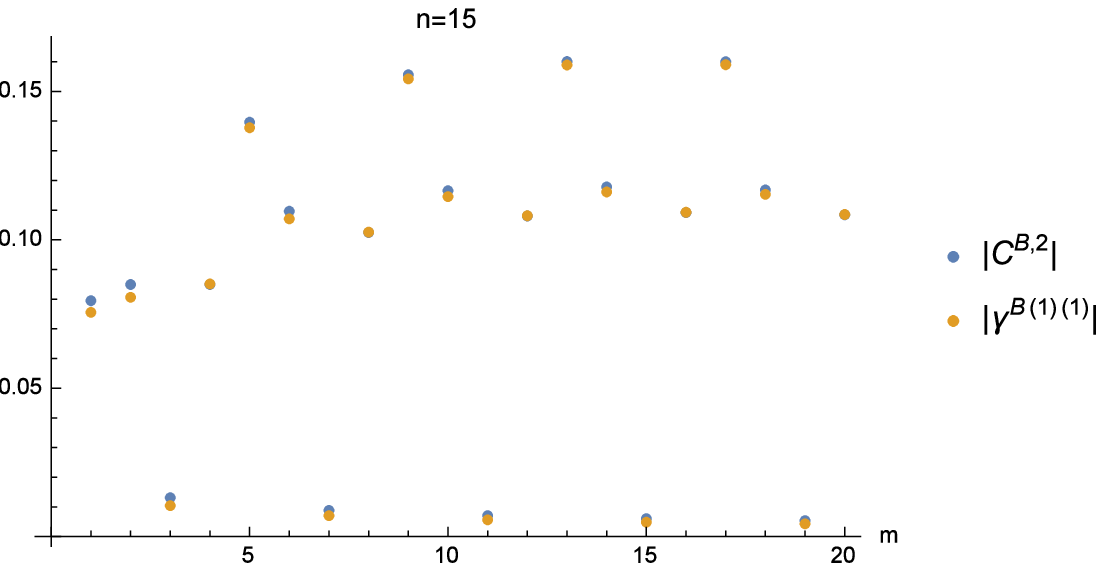}
\caption{Figure 1: Here we plot numerical approximations for both $|C^{B,2}_{mn,++--}|$ and $|\g^{B(1)(1)}_{mn}|$ with $n=10$ (left) and $n=15$ (right) at $\Delta w = {i \pi\over 2}$, for various values of $m$. We see good agreement between the two numerical quantities for each case, as expected.  The small discrepancies for lower $n$ arise from the fact that we have truncated the two infinite sums for some $p,p'>> m,n$.  At low $n$ the convergence takes longer and thus the truncation produces larger discrepancies.}
\label{figure one}
\end{figure}
We can also show that the two twist wick contraction term is related to the complex conjugate of the two twist $\g^{B}$. To do this we again write the two twist wick contraction term:
\bea
C^{B,2}_{mn,++--} &=& {}^{(1)}\langle 0 |  {}^{(2)}\langle 0 | \s_{2}^{+}(w_2)\s_{2}^{+}(w_1)\a_{++,-m}^{(1)}\a_{--,-n}^{(1)}|0\rangle^{(1)}|0\rangle^{(2)}\nn
&=&\left({}^{(1)}\langle 0 |  {}^{(2)}\langle 0 |\a_{++,n}^{(1)}\a_{--,m}^{(1)} \s_{2}^{+}(w_1)\s_{2}^{+}(w_2)|0\rangle^{(1)}|0\rangle^{(2)}\right)^{\dagger}\nn
&=&-mn\left(\g^{B(1)(1)}_{mn,w_{1}\leftrightarrow w_2}\right)^{*}\nn
&=&-mn\g^{B(1)(1)}_{mn}\nn
\label{two twist wick gamma relation}
\eea
where the fourth equality arises because of the reality of $\g^{B(1)(1)}_{mn}$ and the invariance of $a$ and $b$ under the interchange $w_{1}\leftrightarrow w_2$, both conditions of which are a result of our definition of $z_{1}$ and $z_{2}$. In Figure (\ref{figure one}) we present a scatter plot of (\ref{two twist wick final}) and (\ref{two twist wick gamma relation}) where $\g^{B}$ was computed in \cite{acm1}, $f^{B}$ was computed in \cite{acm2}, and $\g^{B(1)(1)}$ was computed in \cite{chm1}. 

Following the same procedure as for the bosons one can compute the fermion two twist wick contraction term, but there are additional complications one must be aware of. The conjugate relations used in the wick contraction computation change nontrivially for fermions because of the fermion modes actually notice the change in charge of the twist operator. Essentially, the conjugate relation is the $\g^{F}$ for a state, $|\chi'\rangle$ built from negative twist operators.  This state has not been computed in any previous work, and we will not do so here. However, the relation does provide an easier way for computing the fermion two twist wick contraction term if necessary.

\section{Proof of $G^{+}_{\dot{A},0}\s_{2}^{+}(w_{0}) = 0$}\label{GPlusProof}
Here we prove the relation that a $G^{+}_{\dot{A},0}$ acting on a single $\s^{+}_{2}$ is equal to zero which is necessary for proving the two twist $\g^{B},\g^{F}$ relations found in \cite{chm1} as well as the $f^{B},f^{F\pm}$ relations given by Equations (\ref{NegativeRelation}) and (\ref{PositiveRelation}) . 
\subsubsection*{Cylinder}
Let us consider the following state on the cylinder:
\bea
{1\over 2 \pi i}\int_{w_0} G^{+}_{\dot{A}}(w)\s_{2}^{+}(w_0) \diff w
\label{cylinder state}
\eea
where $w_0$ is the location of the twist operator. 
\subsubsection*{$z$ plane}
Mapping to the $z$ plane with:
\bea
z=e^{w},\qquad z_0\equiv e^{w_0}
\label{z plane map}
\eea
we find that (\ref{cylinder state}) becomes:
\bea
 z_{0}^{1/2} {1\over 2 \pi i}\int_{z_0} \diff z\, z^{1/2}  G^{+}_{\dot{A}}(z)\s_{2}^{+}(z_0)
 \label{z plane state}
\eea
 where $z_0^{1/2}$ is a Jacobian factor coming from $\s_{2}^{+}(z_0)$.
\subsubsection*{$t$ plane}
Let us map (\ref{z plane map}) to the $t$ plane. Using the single twist map defined in \cite{acm1}
\bea
z = z_0 + t^{2},
\label{t plane map two}
\eea
(\ref{z plane state}) becomes:
\bea
&&\!\!\!\!\!\!\!\!z_{0}^{1/2} {1\over 2 \pi i}\int_{t=0} \diff t\, z^{1/2} \left({\diff z\over \diff t}\right)^{-1/2}G^{+}_{\dot{A}}(t)S^{+}(0)| 0_{NS} \rangle_{t=0} \nn
&& \!\!\!\!\!\!\!\!\quad = 2^{-1/2} z_{0}^{1/2} {1\over 2 \pi i}\int_{t_0} \diff t\, \left(z_0 + t^2\right)^{1/2}t^{-1/2}G^{+}_{\dot{A}}(t)S^{+}(0)| 0_{NS} \rangle_{t=0}\nn
\label{t plane state one}
\eea
\subsubsection*{Spectral Flowing away $S^{+}(0)$}
We now perform a single spectral flow to remove $S^{+}(0)$. The change in the fields are as follows:
\subsubsection*{$\a = -1$ $\text{around}$ $t=0$ }
\bea
S^{+}(0) &\to& 1\nn
G^{+}_{\dot{A}}(t)&\to& G^{+}_{\dot{A}}(t)t^{1/2}
\label{sf t zero}
\eea
Applying (\ref{sf t zero}) to (\ref{t plane state one}) gives:
\bea
2^{-1/2}Cz_{0}^{1/2}{1\over 2 \pi i}\int_{z_0} \diff t\, \left(z_0 +t^{2}\right)^{1/2}G^{+}_{\dot{A}}(t)| 0_{NS} \rangle_{t=0}
\label{t plane state two}
\eea
Let us now expand $\left(z_0 +t^{2}\right)^{1/2}$ around $t=0$:
\bea
 \left(z_0 +t^{2} \right)^{1/2}&=&z_{0}^{1/2}\left(1-z_0^{-1/2} t^{2}\right)^{1/2}\nn
 &=&\sum_{k\geq 0}{}^{1/2}C_{k}z_{0}^{-k/2+1/2}t^{2k}
 \label{expansion}
\eea
Inserting (\ref{expansion}) back into (\ref{t plane state two}) gives:
\bea
2^{-1/2}C\sum_{k\geq 0}{}^{1/2}C_{k}z_{0}^{-k/2+1}{1\over 2 \pi i}\int_{z_0} \diff t\, G^{+}_{\dot{A}}(t)t^{2k} | 0_{NS} \rangle_{t=0}
\label{t plane state three}
\eea
Let us now define $G^{+}_{\dot{A}}$ natural to the $t$ plane at $t=0$:
\bea
\tilde{G}^{+,t\to 0}_{\dot{A},r}&=&{1\over 2 \pi i}\int_{0} \diff t\,  G^{+}_{\dot{A}}(t)t^{r+1/2}, \qquad r \in \mathbb{Z} + 1/2
\label{t plane G modes}
\eea
Rewriting (\ref{t plane state three}) in terms of (\ref{t plane G modes}) gives:
\bea
&&2^{-1/2}C \sum_{k\geq 0}{}^{1/2}C_{k}z_{0}^{-k/2+1} \tilde{G}^{+,t\to 0}_{\dot{A},2k-1/2}| 0_{NS} \rangle_{t=0}
\label{t plane state five}
\eea
It is clear that (\ref{t plane state five}) vanishes when $k\geq 1$:
\bea
\tilde{G}_{\dot{A},k-1/2}^{+,t\to 0}|0_{NS}\rangle_{t=0} = 0,\qquad\qquad k\geq 1
\eea
Let us check the case where $k=0$. Only writing the $G^{+}_{\dot{A}}$ part of (\ref{t plane state five}) we have:
\bea
\tilde{G}_{\dot{A},-1/2}^{+,t\to 0}|0_{NS}\rangle_{t=0}
\label{G on NS}
\eea
This result also vanishes but let us explicitly show this. Splitting our $\tilde{G}^{+}_{\dot{A}}$ into bosons and fermions, a general mode can be written as:
\bea
\tilde{G}^{+,t\to 0}_{\dot{A},r} =-i\sum_{p\in \mathbb{Z}} \tilde{d}^{+A,t\to t_0}_{r-p}\a_{A\dot{A},p}^{t \to 0}
\eea
Applying this to (\ref{G on NS}) we get:
\bea
-i\sum_{p\in \mathbb{Z}} \tilde{d}^{+A,t\to 0}_{1/2-p}\a_{A\dot{A},p}^{t \to 0}|0_{NS}\rangle_{t=0} &=& -i\sum_{p > 0}\tilde{d}^{+A,t\to 0}_{-p+1/2}\a_{A\dot{A},p}^{t \to 0}|0_{NS}\rangle_{t=0}\nn
&&- i\sum_{p > 0}\a_{A\dot{A},-p}^{t \to 0}\tilde{d}^{+A,t\to 0}_{p+1/2}|0_{NS}\rangle_{t=0}\nn
&& -i \tilde{d}^{+A,t\to 0}_{1/2 }\a_{A\dot{A},0}^{t \to 0}|0_{NS}\rangle_{t=0}\nn
&=&0
\eea
Therefore, we have proven that for $k\geq 0$:
\bea
&&\!\!\!\!\!\! 2^{-1/2}C\sum_{k\geq 0}{}^{1/2}C_{k}z_{0}^{-k/2+1} \tilde{G}^{+,t\to 0}_{\dot{A},2k-1/2}| 0_{NS} \rangle_{t=0}=0
\eea
which of course implies that on the cylinder:
\bea
{1\over 2 \pi i}\int_{w_0}G^{+}_{\dot{A}}(w)\s_{2}^{+}(w_0) \diff w =0,
\eea
which is what we wanted to show.

\end{document}